\documentclass{article}
\usepackage[margin=1in]{geometry}
\usepackage{amsmath}
\usepackage{epsfig}
\usepackage{tcolorbox}
\numberwithin{equation}{section} % amsmath

\newcommand{\beq}{\begin{equation}}
\newcommand{\eeq}{\end{equation}}
\newcommand{\beqa}{\begin{eqnarray}}
\newcommand{\eeqa}{\end{eqnarray}}
\newcommand{\bdm}{\begin{displaymath}}
\newcommand{\edm}{\end{displaymath}}
\newcommand{\lslash}[1]{#1\llap/}

\newcommand{\Eq}[1]{Eq.\ (\ref{#1})}
\newcommand{\Eqs}[2]{Eqs.\ (\ref{#1}) and (\ref{#2})}

\newcommand{\Eqsss}[4]{Eqs.\ (\ref{#1}), (\ref{#2}), (\ref{#3}) and (\ref{#4})}
\newcommand{\Eqsto}[2]{Eqs.\ (\ref{#1})-(\ref{#2})}
\newcommand{\xRef}[1]{Ref.\ \cite{#1}}
\newcommand{\xRefs}[2]{Refs.\ \cite{#1} and \cite{#2}}

\newcommand{\Fig}[1]{Fig.\ \ref{#1}}
\newcommand{\Table}[1]{Table\ \ref{#1}}
\newcommand{\Section}[1]{Section\ \ref{#1}}

\newcommand{\fEp}{f}
\newcommand{\fEpp}{f^\prime}
\newcommand{\fbEp}{\bar f}
\newcommand{\fbEpp}{\bar f^\prime}
\newcommand{\fnu}{f^\prime_\nu}
\newcommand{\fnub}{\bar f^\prime_\nu}
\newcommand{\Dp}{\frac{d^4p}{(2\pi)^4}}
\newcommand{\Dpp}{\frac{d^4p^\prime}{(2\pi)^4}}
\title{
  Taming the pinch singularities in the two-loop
  neutrino self-energy in a medium
}
  
\author{Jos\'e F. Nieves\footnote{nieves@ltp.uprrp.edu}\\
  Laboratory of Theoretical Physics, Department of Physics\\
  University of Puerto Rico, R\'{\i}o Piedras, Puerto Rico 00936
  \and\\[12pt]
  Sarira Sahu\footnote{sarira@nucleares.unam.mx}\\
  Instituto de Ciencias Nucleares\\
  Universidad Nacional Aut\'onoma de Mexico\\
  Circuito Exterior, C. U.\\
  A. Postal 70-543, 04510 Mexico DF, Mexico\\
}
\date{}
\begin{document}
\maketitle

\begin{abstract}
  We consider the calculation of the thermal self-energy of a neutrino
  that propagates in a medium composed of fermions and scalars interacting
  via a Yukawa-type coupling, in the case that the neutrino energy
  is much larger than the fermion and scalar masses, as well as the temperature
  and chemical potentials of the background. In this kinematic regime
  the one-loop contribution to the imaginary part of the self-energy
  is negligible. We consider the two-loop contribution
  and we encounter the so-called pinch singularities which are
  known to arise in higher loop self-energy calculations in
  Thermal Field Theory. With a judicious use of the
  properties and parametrizations of the thermal propagators
  the singularities are treated effectively and actually disappear.
  From the imaginary part of the self-energy, we obtain a precise formula
  for the damping matrix expressed in terms of integrals over the background
  particle distributions. The formulas predict a specific dependence
  of the damping terms on the neutrino energy, depending on the background
  conditions. For guidance in estimating the effects in specific contexts,
  we compute the damping terms for several limiting cases
  of the momentum distribution functions of the background particles.
  We discuss briefly the connection between the results of our calculations
  for the damping matrix and the decoherence effects described in terms
  of the Lindblad equation. 
\end{abstract}

\section{Introduction and Summary}
\label{sec:introduction}

In several models and extensions of the standard electro-weak theory
the neutrinos interact with scalar particles ($\phi$)
and fermions ($f$) via a coupling of the form
\beq
\label{Lint}
L_{int} = \sum_a g_a \bar f_R\nu_{La}\phi + h.c\,.
\eeq
For definiteness we are assuming the presence of only one $f$ and $\phi$,
while the indices $a,b,c,...$ label the neutrino flavors.
Those interactions produce nonstandard contributions to the neutrino index
of refraction and effective potential when the neutrino propagates
in a background of those particles. 
Couplings of this form have been considered recently in the context of
Dark Matter-neutrino interactions\cite{Mangano:2006mp,Binder:2016pnr,
Primulando:2017kxf,Campo:2017nwh,Brune:2018sab,Franarin:2018gfk,Dev:2019anc,
Pandey:2018wvh,Karmakar:2018fno}.
Similar effects occur due to neutrino-neutrino-scalar interactions of the
form $\bar\nu^c_{Rb}\nu_{La}\phi$ when a neutrino propagates in a neutrino
background. This can occur in the environment of a supernova,
where the neutrino-neutrino interactions lead to the collective
neutrino oscillations and related
phenomena (see for example \xRefs{Duan:2010bg}{Chakraborty:2016yeg}
and the works cited therein), and it can also occur in the hot
plasma of the Early Universe before the neutrinos
decouple\cite{Wong:2002fa,Mangano:2006ar}.

In previous works we have presented various calculations related to
the propagation of neutrinos in that kind of
background\cite{nsnuphireal,nsnuphidamp,nsnuphidec}.
In \xRef{nsnuphireal} we considered the real part of the self-energy
of a neutrino that propagates in a medium consisting of fermions and scalars,
with a coupling of the form given in \Eq{Lint}.
We calculated the real part (or more precisely
the dispersive part) of the neutrino thermal self-energy, denoted by
$\Sigma_{r}$, from which the dispersion relation and effective potential
are determined. Those interactions can also induce processes such as  
$\nu + \phi \leftrightarrow f$ and $\nu + \bar f\leftrightarrow \bar\phi$,
depending on the kinematic conditions, that produce damping terms
in the neutrino dispersion relation and index of refraction. Thus
in \xRef{nsnuphidamp}, we continued our work
to calculate the imaginary part (or more precisely the absorptive part)
of the neutrino thermal self-energy, denoted by $\Sigma_i$,
in a scalar and fermion background due to the $\bar f_R\nu_L\phi$ interaction.
From $\Sigma_i$ the corresponding contribution to the damping matrix $\Gamma$
in the dispersion relation were obtained. The calculations in
\xRef{nsnuphidamp} were based on the one-loop diagram for the
neutrino self-energy.

In \xRef{nsnuphidec} we noted that those couplings can induce
decoherence effects, of the form discussed in recent works
\cite{Coloma:2018idr,Carpio:2017nui,Oliveira:2014jsa,
Fogli:2007tx,Capolupo:2018hrp},
due to the neutrino non-forward scattering process
$\nu_a + x \rightarrow \nu_b + x$, where $x = f,\phi$.
As observed in \xRef{nsnuphidec}, the
contribution to $\Gamma$ due to these processes
can be determined from the two-loop calculation of $\Sigma_i$.
Thus, in that reference we performed the two-loop calculation
of $\Sigma_i$ and $\Gamma$ or the case in which the background contains only
the fermions $f$, assuming that the $\phi$ particle is heavy enough and the
conditions are such that there are no $\phi$ particles in the background.
Under those conditions, the two-loop contribution to $\Gamma$ is the relevant
one since the two-body processes that contribute in one-loop
are kinematically forbidden.

The present work is a continuation of that previous work. Here we consider
the situation in which both $f$ and $\phi$ may be present in the background.
We are particularly interested in the kinematic regime
\beq
\label{eq:kappacondition}
\kappa > m_\phi, m_f, T\,,
\eeq
where $\kappa$ is the neutrino momentum and $T$ the background temperature.
That is, both $f$ and $\phi$ are relatively light compared to the neutrino
energy. We refer to this as the \emph{light background}.
It is the kinematic regime that is relevant in the context of the
possible existence of light scalars as dark-matter and the
effects they may have on neutrino experiments, that has been explored
in the recent literature\cite{choietal,devetal}.
The results can be useful also for the studies of the environmental
decoherence effects in long baseline neutrino oscillation experiments that have
been carried out recently\cite{bakhtietal,coelhomann,gomesetal,degouveaetal}.
Again, in this kinematic regime the two-body processes
that contribute in one-loop are inhibited and the two-loop contribution
is the relevant one.

Apart from the relevance for the applications already mentioned,
from a calculational point of view the present calculation has a technical
merit. There is one important technical issue that shows up in the kinematic
regime we are considering in the present case and those we considered
previously. The two-loop diagrams for the self-energy, from which
the damping matrix is determined, suffer from the so-called
\emph{pinch-singularities}\cite{pinch}. These arise from the fact that in the
present case some of the diagrams contain a product of two thermal propagators
with the same momentum. Since the thermal propagators involve the on-shell
delta functions, such products are ill-defined. As we show, by a judicious
use of the properties and parametrizations of the thermal propagators,
the expressions for the diagrams can be rearranged such that the
pinch singularities are absent in the final expressions, allowing
a straightforward evaluation of the self-energy and whence the damping
terms. While the conventional wisdom is that indeed such singularities
actually disappear, our calculations provide an explicit proof
of that fact in a concrete and non-trivial example that can be generalized
to other calculations.

The final results are well defined formulas for
the damping terms in the neutrino dispersion relation (or effective potential)
in terms of the model parameters (i.e., couplings $g_a$, masses $m_{f,\phi}$)
and the enviromental parameters (e.g. temperature). In practical applications,
the possible values of all the parameters involved
vary significantly depending on the context, i.e., astrophysical, cosmological
or neutrino oscillations. For example, in the context of a
supernova (such as SN1987A) the neutrino interactions with $\phi$
as a cold dark matter candidate can have effects on the observed neutrino flux
for $m_{\phi}, T \sim \mbox{a few}$ MeV\cite{Mangano:2006mp}. But these,
and similar considerations in other contexts, depend on the particle
physics model as well. Thus, for example, while we concentrate here
on the calculation involving the $L_{int}$ interaction
term, in isolation from the Standard Model interactions, the two-loop diagrams
in some models may involve the standard particles and/or other
non-standard gauge boson interactions as well.
Nevertheless, subject to the limitation of the
\emph{light background} condition stated above, the particular results
we obtain for the damping terms can be used in the context of many such
models and conditions, and in fact the method to treat the pinch singularities
is applicable to those more general cases as well.

In summary, our plan is as follows. In \Section{sec:preliminaries} we
summarize the framework in which we carry out the calculations.
There we explain that, while the effective potential is determined
from the one-loop diagram for the self-energy, in the kinematic
regime we consider [\Eq{eq:kappacondition}] the damping is determined
form the two-loop diagrams. In \Section{sec:veff} we calculate
the dispersive part of the self-energy and determine the effective potential.
In \Section{sec:twoloop} we consider the calculation of the two-loop
contribution to the absorptive part of the self-energy,
from which the damping matrix is determined. There we indicate
the problem of the pinch singularities, and present
our treatment to resolve it. The net result, summarized in
\Section{subsec-twoloop-pinch-summary}, is the set of
formulas for the two-loop contributions to the absorptive part of the
self-energy, free from the singularities. In \Section{sec:purephibackground}
we evaluate explicitly the corresponding expressions for the damping matrix
in the case of a scalar background under various conditions, and indicate
the path to generalize such calculations to consider more complicated
backgrounds. There we also discuss briefly the connection between the damping
matrix thus determined and the decoherence described in terms of the
Lindblad equation. Finally \Section{sec:conclusions} has our conclusions.
\section{Preliminaries - effective potential and the damping matrix}
\label{sec:preliminaries}

To be self-contained we summarize the following material borrowing
from \xRef{nsnuphidec}. We denote by $k^\mu$ the momentum four-vector
of the propagating neutrino, and as usual we denote by $u^\mu$ the velocity
four-vector of the background medium. In the background medium's own
rest frame, it takes the form
\beq
\label{restframe}
u^\mu = (1,\vec 0)\,,
\eeq
and in this frame we write
\beq
\label{krestframe}
k^\mu = (\omega,\vec\kappa)\,.
\eeq
Since we are considering only one background medium, it can be taken
to be at rest and therefore we adopt \Eqs{restframe}{krestframe}
throughout.

Let us consider first the case of one neutrino propagating in the medium.
The dispersion relation and the spinor of
the propagating mode are determined by solving the equation
\beq
\label{eveq}
\left(\lslash{k} - \Sigma_{eff}\right)\psi_L(k) = 0\,,
\eeq
where $\Sigma_{eff}$ is the neutrino thermal self-energy.
It can be decomposed in the form
\beq
\label{Sigmaridecomp}
\Sigma_{eff} = \Sigma_r + i\Sigma_i\,,
\eeq
where $\Sigma_{r,i}$ are the dispersive and absorptive parts,
\beqa
\Sigma_r & = & \frac{1}{2}\left(\Sigma_{eff} + \bar\Sigma_{eff}\right)\,,
\nonumber\\
\Sigma_i & = & \frac{1}{2i}\left(\Sigma_{eff} - \bar\Sigma_{eff}\right)\,,
\eeqa
respectively, with
\beq
\bar\Sigma_{eff} = \gamma^0\Sigma^\dagger_{eff}\gamma^0 \,.
\eeq
In the context of thermal field theory $\Sigma_r$
is given in terms of the $11$ element of the thermal self-energy matrix by
\beq
\label{Sigmardef1}
\Sigma_r = \Sigma_{11r} \equiv
\frac{1}{2}(\Sigma_{11} + \overline\Sigma_{11})\,.
\eeq
On the other hand, $\Sigma_i$ is more conveniently determined in terms
of the $12$ element of the neutrino thermal self-energy matrix by
the formula
\beq
\label{Sigmaidef1}
\Sigma_i = \frac{\Sigma_{12}}{2i n_F(x_\nu)}\,.
\eeq
Here
\beq
\label{nF}
n_F(z) = \frac{1}{e^z + 1}\,,
\eeq
is the fermion distribution function, written in terms of a dummy variable $z$,
and the variable $x_\nu(k)$ is given by
\beq
x_\nu(k) = \beta k\cdot u - \alpha_\nu\,,
\eeq
where $T = 1/\beta$ is the temperature and $\alpha_\nu$
is the neutrino chemical potential.

The chirality of the neutrino interactions imply that\footnote{%
In a strict sense this is correct in the massless neutrino
limit, which is valid in practice in the approximation that
the neutrino mass is neglected in the calculation of the
relevant diagrams for the self-energy.}
\beq
\label{SigmaV}
\Sigma_{eff} = V^\mu\gamma_\mu L\,.
\eeq
Here and below we use the notation $L$ and $R$
for the left and right chiral projection matrices
$L,R = \frac{1}{2}(1 \mp \gamma_5)$, respectively.
Corresponding to the decomposition in \Eq{Sigmaridecomp} we also write
\beq
\label{Vri}
V^\mu = V^\mu_r + iV^\mu_i\,.
\eeq
and
\beq
\label{SigmariVri}
\Sigma_{r,i} = V^\mu_{r,i}\gamma_\mu L\,.
\eeq
In general $V^\mu_{r,i}$ are functions of $\omega$ and $\vec\kappa$.
We omit those arguments ordinarily but we will restore them when needed.

Writing the neutrino and antineutrino dispersion relations in the form
\beq
\label{disprelform}
\omega^{(\nu,\bar\nu)}(\kappa) = \omega^{(\nu,\bar\nu)}_r(\kappa) -
\frac{i\gamma^{(\nu,\bar\nu)}(\kappa)}{2}\,,
\eeq
the solution of \Eq{eveq} gives
\beq
\label{nudisprelreal}
\omega^{(\nu,\bar\nu)}_r = \kappa + V^{(\nu,\bar\nu)}_{eff}
\eeq
where $V^{(\nu,\bar\nu)}_{eff}$ are the effective potentials
\beqa
\label{Veff}
V^{(\nu)}_{eff} & = & n\cdot V_r(\kappa,\vec\kappa) =
 V^0_r(\kappa,\vec\kappa) - \hat\kappa\cdot\vec V_r(\kappa,\vec\kappa)
\,,\nonumber\\
V^{(\bar\nu)}_{eff} & = & -n\cdot V_r(-\kappa,-\vec\kappa) =
 -V^0_r(-\kappa,-\vec\kappa) + \hat\kappa\cdot\vec V_r(-\kappa,-\vec\kappa)\,,
\eeqa
with
\beq
\label{nmu}
n^\mu = (1,\hat\kappa)\,.
\eeq
On the other hand, for the imaginary part,
\beqa
\label{nudisprelimg}
-\frac{\gamma^{(\nu)}(\vec\kappa)}{2} & = & 
\frac{n\cdot V_i(\kappa,\vec\kappa)}
{1 - n\cdot\left.
\frac{\partial V_r(\omega,\vec\kappa)}{\partial\omega}
\right|_{\omega = \kappa}}\,,\nonumber\\
-\frac{\gamma^{(\bar\nu)}(\vec\kappa)}{2} & = & 
\frac{n\cdot V_i(-\kappa,-\vec\kappa)}
{1 - n\cdot\left.
\frac{\partial V_r(\omega,-\vec\kappa)}{\partial\omega}
\right|_{\omega = -\kappa}}\,,
\eeqa
where $n^\mu$ is defined in \Eq{nmu}. We will retain only the dominant
contribution to $n\cdot V_i$ in the numerator,
which in our case is the two-loop term as we argue below.
Then to leading order the formulas in \Eq{nudisprelimg} reduce to
\beqa
\label{nudisprelimg-simple}
-\frac{\gamma^{(\nu)}(\vec\kappa)}{2} & = & 
n\cdot V_i(\kappa,\vec\kappa)\,,\nonumber\\
-\frac{\gamma^{(\bar\nu)}(\vec\kappa)}{2} & = & 
n\cdot V_i(-\kappa,-\vec\kappa)\,.
\eeqa
neglecting the correction due to the
$n\cdot\partial V_r(\omega,\vec\kappa)/\partial\omega$ term
in the denominator.

In the case of various neutrino flavors, the vector $V^\mu$ in 
\Eq{SigmaV} is a matrix in the neutrino flavor space.
As shown in \xRef{nsnuphidec}, the generalization of the above
discussion is that the dispersion
relations of the propagating modes are determined by solving
the following eigenvalue equation in flavor-space,
\beq
\left(H_r - i\frac{\Gamma}{2}\right)\xi = \omega\xi\,,
\eeq
where $H_r$ and $\Gamma$ are Hermitian matrices in flavor space given by
\beqa
\label{HrGammaVrelation}
H_r & = & \left\{\begin{array}{ll}
\kappa + n\cdot V_r(\kappa,\vec\kappa) & (\nu)\\
\kappa - n\cdot V^\ast_r(-\kappa,-\vec\kappa) & (\bar\nu)\,,
\end{array}\right.\nonumber\\[12pt]
-\frac{1}{2}\Gamma & = & \left\{\begin{array}{ll}
n\cdot V_i(\kappa,\vec\kappa) & (\nu)\\
n\cdot V^\ast_i(-\kappa,-\vec\kappa) & (\bar\nu)\,.
\end{array}\right.
\eeqa
In coordinate space, this translates to the evolution equation
\beq
\label{eveqt}
i\partial_t\xi(t) = \left(H_r - i\frac{\Gamma}{2}\right)\xi(t)\,.
\eeq
We refer to $\Gamma$ as the damping matrix,
and to its elements as the damping terms.
Our purpose in this work is to determine the contribution to $H_r$,
and specially $\Gamma$, due to the presence of the \emph{light background}.

\begin{figure}
\begin{center}
\epsfig{file=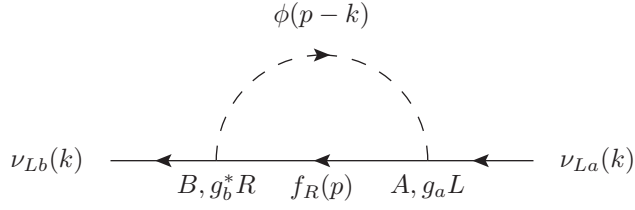,bbllx=189,bblly=351,bburx=423,bbury=426}
\end{center}
\caption[]{
  One-loop diagram for the neutrino self-energy matrix
  $(\Sigma_{AB})_{ba}$
  in a background of fermions $f$ and scalars $\phi$.
  $a,b$ are neutrino flavor indices while
  $A$ and $B$ label the thermal vertices, that can take the values 1 or 2.
  \label{fig:oneloop}
}
\end{figure}

The lowest order diagram is shown in \Fig{fig:oneloop}. From
that diagram we obtain $\Sigma_{11}$ and $\Sigma_{12}$, and whence
the dispersive and absorptive parts $\Sigma_{r,i}$ by means
of \Eqs{Sigmardef1}{Sigmaidef1}. The corresponding one-loop
contribution to $V_r$ and $\Gamma$ are then obtained from
\Eq{HrGammaVrelation}.
However, as in the cases discussed in \xRefs{nsnuphidec}{nsnudecsm},
the one-loop contribution to $\Sigma_{12}$ is negligible in this case also.
The reason is that such contributions arise from 
the two-body neutrino processes such as $\nu + \phi\leftrightarrow f$,
which are inhibited by the kinematics in the regime we are considering
[i.e, \Eq{eq:kappacondition}].

\begin{figure}
\begin{center}
\epsfig{file=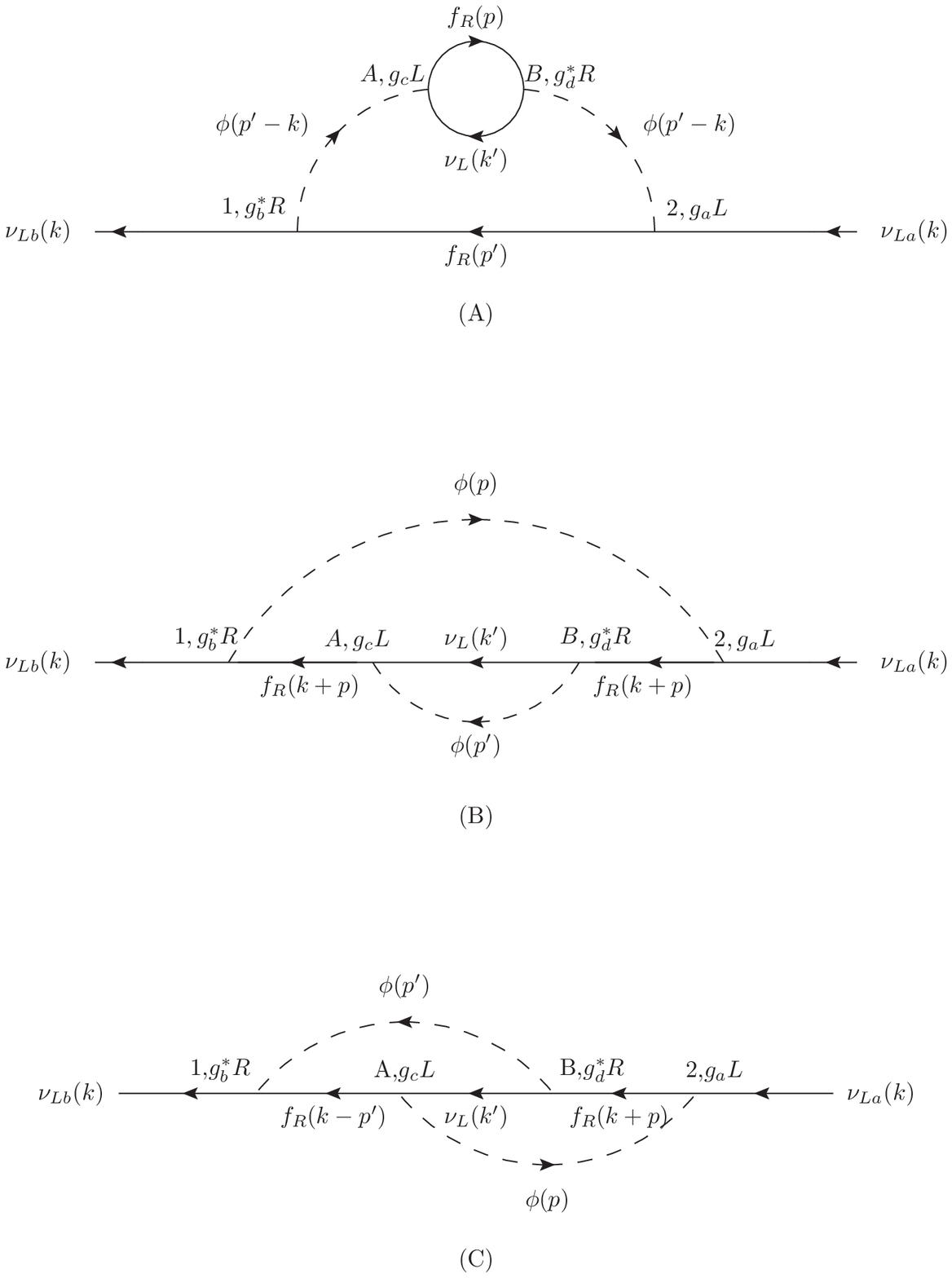,bbllx=109,bblly=146,bburx=503,bbury=672}
\end{center}
\caption[]{
  Two-loop diagram for the 12 element of the neutrino self-energy matrix
  in a background of fermions $f$ and scalars $\phi$, obtained by
  inserting additional propagators in the one-loop diagram in \Fig{fig:oneloop}.
  The meaning of the labels $A,B$ are the same as in \Fig{fig:oneloop}
  while $a,b,c,d$ are neutrino flavor indices.
  To simplify the notation we are setting $k^\prime = k + p - p^\prime$.
  Diagram C does not exist if $\phi$ is a complex scalar
  field. For a real scalar diagram C must be included.
  \label{fig:twoloop}
}
\end{figure}

To be more specific, the one-loop damping term is due to real processes like
\beqa
\label{oneloopdampingprocesses}
(A) &
\nu + \bar f & \leftrightarrow \bar\phi \quad \mbox{(if $m_\phi > m_f$)}
\nonumber\\
(B)& \nu + \phi & \leftrightarrow f \quad \mbox{(if $m_f > m_\phi$)}\,.
\eeqa
The calculation of the one-loop damping terms for all such conditions
was carried out in \xRef{nsnuphidamp}. Let us consider (A).
As shown in that reference, the damping is maximum for values of the neutrino
momentum
\beqa
\label{eq:dampingmaxcond}
\kappa & \sim & \frac{m^2_\phi}{T} \quad \mbox{(if $T \gg m_f$)}\nonumber\\
\kappa & \sim & \frac{m^2_\phi}{m_f} \quad \mbox{(if $T \ll m_f$)}\,.
\eeqa
Outside of those ranges the damping becomes exponentially small.
Analogous considerations apply to case (B) as well.
As a result, in the kinematic regime we are considering,
the damping matrix is determined by the two-loop diagrams
for $\Sigma_{12}$ shown in \Fig{fig:twoloop}.

In summary, the neutrino and antineutrino effective potential
is given by Eq. (2.16), and the damping by Eq. (2.19), where
$V^\mu_r$ is determined from the calculation of
$\Sigma_{11}$ using the one-loop diagram in \Fig{fig:oneloop},
while $V^\mu_i$ is determined from $\Sigma_{12}$ calculated from the
two-loop diagrams in \Fig{fig:twoloop}. As already stated,
in writing Eq. (2.19) we are neglecting the correction due to the
$n\cdot\partial V_r(\omega,\vec\kappa)/\partial\omega$ in the
overall denominator in Eq (2.18), which corresponds to keep the leading order
of the dominant term.
As we have emphasized, the expressions corresponding to the diagrams
in \Fig{fig:twoloop} suffer from the pinch singularities. 
After handling the singularities, $\Sigma_{12}$ is expressed
in terms of integrals over the background particles distribution
functions that can be evaluated in principle once the background
conditions are specified. Correspondingly, the final formula for
the damping matrix that we determine by means of \Eq{HrGammaVrelation} 
is expressed in terms of integrals over the background particle distribution
functions that we will evaluate explicitly for some illustrative cases.

\section{Effective potential}
\label{sec:veff}

\subsection{Dispersive part $\Sigma_r$}

The contribution of the diagram in \Fig{fig:oneloop}
to the 11 component of the neutrino thermal self-energy matrix
is given by
\beq
\label{Sigma11_1}
-i(\Sigma_{11})_{ba} = \int\,\frac{d^4p}{(2\pi)^4}
i\Delta^{(\phi)}_{11}(p - k)(ig^\ast_b R)iS^{(f)}_{11}(p)(ig_a L) \,.
\eeq
We write the 11 components of the $f$ and $\phi$ thermal propagators
in the form
\beqa
\label{propagators11}
iS^{(f)}_{11}(p) & = & (\lslash{p} + m_f)\left[\frac{i}{p^2 - m^2 + i\epsilon}
  - 2\pi\delta(p^2 - m^2_f)\eta_f(p)\right]\,,\nonumber\\
i\Delta^{(\phi)}_{11}(p) & = & \frac{i}{p^2 - m^2_\phi + i\epsilon}
  + 2\pi\delta(p^2 - m^2_\phi)\eta_\phi(p)\,,
\eeqa
where
\beqa
\label{defeta}
\eta_f(p) & = &
n_F(x_f(p))\theta(p\cdot u) + n_F(-x_f(p))\theta(-p\cdot u)\,,\nonumber\\
\eta_\phi(p) & = &
n_B(x_f(p))\theta(p\cdot u) + n_F(-x_f(p))\theta(-p\cdot u)\,.
\eeqa
with $\theta$ being the unit step function.
Here $n_{F}$ is the fermion momentum distribution function defined in \Eq{nF},
and $n_B$ is the corresponding one for bosons,
\beq
\label{nB}
n_B(z) = \frac{1}{e^z - 1}\,.
\eeq
The variables $x_{f,\phi}$ are given by
\beqa
x_f(p) = \beta p\cdot u - \alpha_f\,,\nonumber\\
x_\phi(p) = \beta p\cdot u - \alpha_\phi\,,
\eeqa
where $\alpha_{f,\phi}$ are the chemical potentials.
Discarding the pure vacuum contribution in \Eq{Sigma11_1} we then have
\beq
\Sigma_r = \Sigma^{(f)}_r + \Sigma^{(\phi)}_r\,,
\eeq
where
\beqa
\label{Sigmarf}
\left(\Sigma^{(f)}_r\right)_{ba} & = & -g^\ast_b g_a\int\frac{d^4p}{(2\pi)^3}
\frac{\lslash{p}L}{(p - k)^2 - m^2_\phi}\delta(p^2 - m^2_f)\eta_f(p)\,,\\
\label{Sigmarphi}
\left(\Sigma^{(\phi)}_r\right)_{ba} & = & g^\ast_b g_a\int\frac{d^4p}{(2\pi)^3}
\frac{(\lslash{p} + \lslash{k})L}{(p + k)^2 - m^2_f}
\delta(p^2 - m^2_\phi)\eta_\phi(p)\,.
\eeqa

\subsection{Effective potential in the \emph{light background}}

For completeness and to make the present work
self-contained, here we consider specifically the case
of the light background in the sense of \Eq{eq:kappacondition}.
In correspondence to \Eq{SigmariVri}, we write the
$\Sigma^{(x)}_r$ ($x = f,\phi$) in the form
\beq
\label{SigmaxV}
\Sigma^{(x)}_r = V^{(x)\mu}_r\gamma_\mu L\,,
\eeq
and therefore
\beq
V^\mu_r = V^{(f)\mu}_r + V^{(\phi)\mu}_r\,.
\eeq
From \Eqs{Sigmarf}{Sigmarphi} we then have
\beqa
\label{Vrf1}
\left(V^{(f)\mu}_r(\omega,\kappa)\right)_{ba} & = &
-(g^\ast_b g_a)\int\frac{d^4p}{(2\pi)^3}
\frac{p_\mu}{(p - k)^2 - m^2_\phi}
\delta(p^2 - m^2_f)\eta_f(p)\,,\\
\label{Vrphi1}
\left(V^{(\phi)\mu}_r(\omega,\kappa)\right)_{ba} & = &
(g^\ast_b g_a)\int\frac{d^4p}{(2\pi)^3}
\frac{(p^\mu + k^\mu)}{(p + k)^2 - m^2_f}
\delta(p^2 - m^2_\phi)\eta_\phi(p)\,.
\eeqa
Under the conditions that we are considering [i.e., \Eq{eq:kappacondition}],
we can make the replacement
\beqa
\label{lightbackgroundexpansion}
\frac{1}{(p + k)^2 - m^2_f} & \rightarrow &
\frac{1}{2p\cdot k + k^2}\,,\nonumber\\
\frac{1}{(p - k)^2 - m^2_\phi} & \rightarrow &
\frac{-1}{2p\cdot k - k^2}\,,
\eeqa
in \Eqs{Vrf1}{Vrphi1}. Furthermore, since the effective potential is
defined by setting $\omega = \kappa$ [i.e., \Eq{HrGammaVrelation}],
we can set
\beq
k^\mu = \kappa n^\mu\,,
\eeq
where $n^\mu$ is defined in \Eq{nmu}. Thus,
\beqa
\label{Vrf2}
\left(V^{(f)\mu}_r(\kappa,\kappa)\right)_{ba} & = &
(g^\ast_b g_a)\int\frac{d^4p}{(2\pi)^3}
\frac{p_\mu}{2p\cdot k}\delta(p^2 - m^2_f)\eta_f(p)\,,\\
\label{Vrphi2}
\left(V^{(\phi)\mu}_r(\kappa,\kappa)\right)_{ba} & = &
(g^\ast_b g_a)\int\frac{d^4p}{(2\pi)^3}
\frac{(p^\mu + \kappa n^\mu)}{2p\cdot k}\delta(p^2 - m^2_\phi)\eta_\phi(p)\,.
\eeqa
Carrying out the integral over $p^0$ with the help of the delta function,
we then have
\beqa
\label{Vx2}
\left(V^{(f)\mu}_r(\kappa,\vec \kappa)\right)_{ba} & = &
\frac{g^\ast_b g_a}{2\kappa}\int\frac{d^3p}{(2\pi)^3 2E_f}
\frac{p^{\mu}}{p\cdot n}
\left[f_f(E_f) + f_{\bar f}(E_f)\right]\,,\nonumber\\
\left(V^{(\phi)\mu}_r(\kappa,\vec \kappa)\right)_{ba} & = &
\frac{g^\ast_b g_a}{2\kappa}\int\frac{d^3p}{(2\pi)^3 2E_\phi}
\frac{1}{p\cdot n}\left\{
p^\mu\left[f_\phi(E_\phi) + f_{\bar\phi}(E_\phi)\right]\right.\nonumber\\
&&\mbox{} + \left.
\kappa n^\mu\left[f_\phi(E_\phi) - f_{\bar\phi}(E_\phi)\right]\right\}\,,
\eeqa
and
\beq
V^{(x)\mu}_r(-\kappa,-\vec \kappa) =
\left.V^{(x)\mu}_r(\kappa,\vec \kappa)\right|_{n^\mu\rightarrow -n^\mu}\,.
\eeq
In \Eq{Vx2} we have introduced the $f$ and $\phi$ momentum
distribution functions (in the rest frame of the medium)
\beqa
\label{thermaldist}
f_{f,\bar f}(E_f) & = &
\frac{1}{e^{\beta E_f \mp \alpha_f} + 1}\,,\nonumber\\
f_{\phi,\bar\phi}(E_\phi) & = & 
\frac{1}{e^{\beta E_\phi \mp \alpha_\phi} -1}\,,
\eeqa
and it is understood that in each integral,
\beq
p^\mu = (E_x,\vec p) \qquad (x = f,\phi)\,,
\eeq
with
\beq
E_x = \sqrt{{\vec p}{\,^2} + m^2_x}\,.
\eeq
Thus remembering that $n^2 = 0$,
\beq
\label{Vx3}
\left(n\cdot V^{(x)}(\kappa,\vec \kappa)\right)_{ba} =
-\left(n\cdot V^{(x)}(-\kappa,-\vec \kappa)\right)_{ba} =
\frac{g^\ast_b g_a}{2\kappa}
\int\frac{d^3p}{(2\pi)^3 2E_x}
\left[f_x(E_x) + f_{\bar x}(E_x)\right]\,.
\eeq
From \Eq{Veff}, we then have
\beq
\label{Vefffinal}
\left(V^{(\nu)}_{eff}(\vec\kappa)\right)_{ba} =
\left(V^{(\bar\nu)}_{eff}(\vec\kappa)\right)_{ba} =
(V_f)_{ba} + (V_\phi)_{ba}\,,
\eeq
where ($x = f,\phi$)
\beq
\label{Vfphi}
(V_x)_{ba} = \frac{g^\ast_b g_a}{2\kappa}J_x\,,
\eeq
and we have defined
\beq
\label{Jfphi}
J_x \equiv \int\frac{d^3p}{(2\pi)^3 2E_x}
\left[f_x(E_x) + f_{\bar x}(E_x)\right]\,.
\eeq
\Eqsto{Vefffinal}{Jfphi} reveal a number of
differences in contrast with the Wolfenstein term that gives
standard matter contribution to the effective potential\cite{wterm}.
The effective potential in this case is momentum-dependent, proportional
to $1/\kappa$, has the same sign for neutrinos and antineutrinos,
and does not vanish in a particle-antiparticle symmetric background.
In fact, as is well known, for practical purposes
the parameters $g^\ast_b g_a J_f$ and
$g^\ast_b g_a J_\phi$ act as contributions to the vacuum squared mass matrix,
with the same value (and sign) for neutrinos and
antineutrinos (see e.g., \xRef{ge-parke}).

It is a simple matter to evaluate $J_{x}$ for different conditions of the
background. For example, and for reference purposes,
in the non-relativistic (NR) limit, or in the ultra-relativistic (UR) limit
and zero chemical potential,
\beq
\label{Jphinr-er}
J_\phi = \left\{
\begin{array}{ll}
\frac{1}{2m_\phi}(n_\phi + n_{\bar\phi}) & \mbox{(NR)}\\[12pt]
\frac{T^2}{12} & \mbox{(UR)}\,,
\end{array}
\right.
\eeq
and similarly for $J_f$. In \Eq{Jphinr-er}, $n_{\phi,\bar\phi}$
are the total number densities of $\phi$ and $\bar\phi$, respectively, i.e.,
\beq
n_{\phi,\bar\phi} = \int\frac{d^3p}{(2\pi)^3} f_{\phi,\bar\phi}(E_\phi)\,.
\eeq
In \Eq{Jphinr-er} we have assumed that $\phi$ is complex. For a real
$\phi$, in the NR limit
\beq
\label{Jphinr-real}
J_\phi = \frac{n_\phi}{m_\phi} \qquad \mbox{(NR)}\,.
\eeq
In the UR limit the formula in \Eq{Jphinr-er} holds in this case
as well.

\section{Two-loop diagrams - absence of pinch singularities}
\label{sec:twoloop}

We reiterate that we assume $\kappa$ to be high enough so that the conditions
such as those given in \Eq{eq:dampingmaxcond} (or the analogous ones
in case (B)) are satisfied and therefore the one-loop contribution to
the damping matrix, arising from the two-body processes shown in
\Eq{oneloopdampingprocesses} is negligible. As we have already mentioned,
in that case the damping terms arise from the two-loop diagrams for
$\Sigma_{12}$ shown in \Fig{fig:twoloop}, which are obtained by inserting
additional propagators in the one-loop diagram in \Fig{fig:oneloop}.
The corresponding expressions for their contributions to $\Sigma_{12}$ are
\beqa
\label{explicitsigma12A}
-i\left(\Sigma^{(A)}_{12}(k)\right)_{ba} & = &
\sum_{A,B}\sum_{c,d}
\int\Dp \Dpp (i\eta_1 g^\ast_b R)iS^{(f)}_{12}(p^\prime)(i\eta_2 g_a L)
i\Delta^{(\phi)}_{2B}(p^\prime - k)i\Delta^{(\phi)}_{A1}(p^\prime - k)
\nonumber\\
&&\mbox{}\times
(-1)\mbox{Tr}\left\{(i\eta_A g_c L)
\left(iS^{(\nu)}_{AB}(k^\prime)\right)_{cd}
(i\eta_B g^\ast_d R) iS^{(f)}_{BA}(p)\right\}\,,
\eeqa
\beqa
\label{explicitsigma12B}
-i\left(\Sigma^{(B)}_{12}(k)\right)_{ba} & = & \sum_{A,B}\sum_{c,d}
\int\Dp \Dpp 
i\Delta^{(\phi)}_{21}(p)i\Delta^{(\phi)}_{AB}(p^\prime)
(i\eta_1 g^\ast_b R)iS^{(f)}_{1A}(k + p)(i\eta_A g_c L)\nonumber\\
&&\mbox{}\times
\left(iS^{(\nu)}_{AB}(k^\prime)\right)_{cd} (i\eta_B g^\ast_d R)
iS^{(f)}_{B2}(k + p) (i\eta_2 g_a L)\,,
\eeqa
\beqa
\label{explicitsigma12C}
-i\left(\Sigma^{(C)}_{12}(k)\right)_{ba}
& = & \sum_{A,B}\sum_{c,d}
\int\Dp \Dpp 
i\Delta^{(\phi)}_{2A}(p)i\Delta^{(\phi)}_{1B}(p^\prime)
(i\eta_1 g^\ast_b R)iS^{(f)}_{1A}(k - p^\prime)(i\eta_A g_c L)\nonumber\\
&&\mbox{}\times
\left(iS^{(\nu)}_{AB}(k^\prime)\right)_{cd} (i\eta_B g^\ast_d R)
iS^{(f)}_{B2}(k + p) (i\eta_2 g_a L)\,,
\eeqa
where
\beq
\label{kprime}
k^\prime = k + p - p^\prime\,.
\eeq
As indicated in Figs.\ \ref{fig:oneloop} and \ref{fig:twoloop},
the subscripts $A,B$ label the internal thermal vertices, and
each of them can take the values 1 or 2. Correspondingly,
the factors $\eta_{A,B}$ take into account the sign of the coupling
associated with each vertex type, $\eta_1 = 1, \eta_2 = -1$.

Apart from the fact that there are many diagrams because we have to sum over
all the values of the internal thermal vertices $A,B$ in each one,
if we attempt to evaluate the expressions for each literally,
we encounter the pinch singularity problems
in diagrams A and B. Take for example diagram A. Since the $\phi$
propagators have the same momentum, there will be diagrams
in which two delta functions (coming from the thermal part of the propagators)
with the same argument, appear. Something similar happens with the
$f$ propagators also in diagram B, while diagram C
does not have this problem. Therefore, before we can proceed we must
prove that those \emph{pinch} singularities actually disappear.
This is what we do next. The end result is a simplified expression
for each contribution $\Sigma^{(A,B,C)}_{12}$ that we then
evaluate explicitly (in the kinematic regime that we are considering).
For convenience and future reference, the relevant formulas are
summarized in \Section{subsec-twoloop-pinch-summary}.
%
% Pinch singularities
%

\subsection{Diagram A}

We write the contribution from diagram A in \Fig{fig:twoloop},
given in \Eq{explicitsigma12A}, in the form
\beqa
\label{formalsigma12A}
-i\left(\Sigma^{(A)}_{12}(k)\right)_{ba} & = &
\sum_{A,B}\int\Dpp 
(i\eta_1 g^\ast_b R)iS^{(f)}_{12}(p^\prime)(i\eta_2 g_a L)\nonumber\\
&&\mbox{}\times
i\Delta^{(\phi)}_{2B}(p^\prime - k)
\left(-i\pi^{(\phi)}_{BA}(p^\prime - k)\right)
i\Delta^{(\phi)}_{A1}(p^\prime - k)\,,
\eeqa
where
\beq
\label{piAB}
-i\pi^{(\phi)}_{BA}(p^\prime - k) =
(-1)\sum_{c,d}\int\Dp \mbox{Tr}\left\{(i\eta_A g_c L)
\left(iS^{(\nu)}_{AB}(k^\prime)\right)_{cd}
(i\eta_B g^\ast_d R) iS^{(f)}_{BA}(p)\right\}\,.
\eeq
The main point is that in this form we can manipulate the integration
in \Eq{formalsigma12A}, and in particular the pinch singularity, using
the symmetry properties and the parametrization of the scalar propagator
as well as the scalar self-energy. Thus we study (and manipulate)
\beq
X_{21}(q) \equiv \sum_{A,B}
\Delta^{(\phi)}_{2B}(q)\,\pi^{(\phi)}_{BA}(q)\,\Delta^{(\phi)}_{A1}(q)\,,
\eeq
which we write in matrix notation,
\beq
X_{21} = \left(\tilde\Delta^{(\phi)}\tilde\pi^{(\phi)}
\tilde\Delta^{(\phi)}\right)_{21}\,.
\eeq
$\tilde\Delta^{(\phi)}$ and $\tilde\pi^{(\phi)}$
are the scalar thermal propagator and self-energy matrices.
We omit the momentum argument $q$ except when required.

We will use the parametrizations given in Eqs (2.9) and (2.34)
of \xRef{canonical}, namely
\beqa
\label{scalarparametrization}
\tilde\Delta^{(\phi)} & = & U_\phi\left(
\begin{array}{cc}
  \Delta^{(\phi)}_F & 0\\
  0 & -\Delta^{(\phi)\ast}_F
\end{array}
\right)U_\phi
\,,\nonumber\\
\tilde\pi^{(\phi)} & = & U^{-1}_\phi\left(
\begin{array}{cc}
  \pi^{(\phi)} & 0\\
  0 & -\pi^{(\phi)\ast}
\end{array}
\right)U^{-1}_\phi\,,
\eeqa
where $\Delta^{(\phi)}_F$ is the (vacuum) Feynman propagator.
The matrix $U_\phi$ is given by
\beq
\label{Uphi}
U_{\phi} = \frac{1}{\sqrt{1 + \eta_\phi(q)}}\left(
\begin{array}{cc}
1 + \eta_\phi(q) & \eta_\phi(q) + \theta(-q\cdot u)\\
\eta_\phi(q) + \theta(q\cdot u) & 1 + \eta_\phi(q)
\end{array}\right)\,,
\eeq
where $\eta_\phi$ is defined in \Eq{defeta}. Therefore, we have
\beq
\tilde X \equiv \tilde\Delta^{(\phi)}\tilde\pi^{(\phi)}\tilde\Delta^{(\phi)}
= U_\phi\left(
\begin{array}{cc}
  \Delta^{(\phi)}_F\pi^{(\phi)}\Delta^{(\phi)}_F & 0\\
  0 & -\Delta^{(\phi)\ast}_F\pi^{(\phi)\ast}\Delta^{(\phi)\ast}_F
\end{array}
\right)U_\phi\,.
\eeq
The key point is that the problematic terms
$\Delta^{(\phi)}_F\Delta^{(\phi)\ast}_F$ that
give rise to the pinch singularity are actually absent. The quantity
that enters in the expression for $\Sigma^{(A)}_{12}$ is
\beqa
X_{21} & = & U_{\phi\,11} U_{\phi\,21}\left[
  \Delta^{(\phi)}_F\pi^{(\phi)}\Delta^{(\phi)}_F -
  \Delta^{(\phi)\ast}_F \pi^{(\phi)\ast}\Delta^{(\phi)\ast}_F\right]
\nonumber\\
& = & (\eta_\phi(q) + \theta(q\cdot u))
\left[
  \Delta^{(\phi)}_F\pi^{(\phi)}\Delta^{(\phi)}_F -
  \Delta^{(\phi)\ast}_F \pi^{(\phi)\ast}\Delta^{(\phi)\ast}_F\right]\,.
\eeqa
Finally, it should be remembered that $\pi$ can be calculated
by using the formulas
\beqa
\label{pi}
\mbox{Re}\,\pi^{(\phi)} & = & \mbox{Re}\,\pi^{(\phi)}_{11}\,,\nonumber\\
\mbox{Im}\,\pi^{(\phi)} & = &
\frac{i\pi^{(\phi)}_{12}}{2[\eta_\phi(q) + \theta(-q\cdot u)]} =
\frac{i\pi^{(\phi)}_{21}}{2[\eta_\phi(q) + \theta(q\cdot u)]}\,,
\eeqa
where the $\pi^{(\phi)}_{AB}$ are given by the expressions in \Eq{piAB}.

Thus in summary the procedure we need to follow is the following:

\begin{enumerate}
\item Compute $\mbox{Re}\,\pi^{(\phi)}_{11}(p^\prime - k)$ and
  $\pi^{(\phi)}_{12}(p^\prime - k)$ using \Eq{piAB}, and from this
  determine $\pi$ using \Eq{pi}.

\item Determine the quantity
  \beq
  \label{X21}
  X_{21}(q) = (\eta_\phi + \theta(q\cdot u))
  \left[\Delta^{(\phi)}_F\pi^{(\phi)}\Delta^{(\phi)}_F -
    \Delta^{(\phi)\ast}_F \pi^{(\phi)\ast}\Delta^{(\phi)\ast}_F\right]\,,
  \eeq
  where $q = p^\prime - k$.

  \item Compute $\Sigma^{(A)}_{12}$ from
    \beqa
    \label{practicalsigma12A}
    -i\left(\Sigma^{(A)}_{12}(k)\right)_{ba} & = &
    \int \Dpp (i\eta_1 g^\ast_b R)iS^{(f)}_{12}(p^\prime)
    (i\eta_2 g_a L)iX_{21}(p^\prime - k)\,.
    \eeqa
    At this point, by means of \Eq{practicalsigma12A},
    these instructions provide a complete and
    consistent procedure for calculating the contribution to the
    neutrino self-energy from diagram A, which is free from the
    pinch singularities.
    However, for our purposes, we can simplify the explicit computation
    as follows.
    
  \item As we argue below, in the kinematic regime we are considering,
    the $\mbox{Re}\,\pi^{(\phi)}$ term in \Eq{X21} contributes
    negligibly, so that we can take $X_{21}$ to be
    \beq
    \label{X21final}
    X_{21} = -\frac{1}{2}\left[
      \Delta^{(\phi)}_F\pi^{(\phi)}_{21}\Delta^{(\phi)}_F +
      \Delta^{(\phi)\ast}_F\pi_{21}\Delta^{(\phi)\ast}_F
      \right]\,.
    \eeq
    The argument concerning the contribution from $\mbox{Re}\,\pi^{(\phi)}$ is
    the following. The term
    $(\Delta^{(\phi)}_F)^2 - (\Delta^{(\phi)\ast}_F)^2$ contributes
    only when the scalar is on-shell, that is
    $(p^\prime - k)^2 \approx m^2_\phi$. Since the external neutrino
    momentum $k$ as well as the fermion momentum $p^\prime$ are on shell
    in \Eq{practicalsigma12A}, the $\mbox{Re}\,\pi^{(\phi)}$ term corresponds
    to a contribution to $\Sigma^{(A)}_{12}$ arising from a process involving
    two-body subproceses $\nu\phi \leftrightarrow f$. Under the kinematic
    conditions we are considering, such sub-processes are suppressed
    by the same (two-body) kinematics that suppress the one-loop contributions
    to $\Sigma^{(\nu)}_{12}$ (and the \emph{raison-de-entre} for considering the
    two-loop contributions) and therefore those terms can be neglected.
    
  \item More specifically, in our case in which we consider
    the high $k$ limit, we take
    \beq
    \Delta^{(\phi)}_F(p^\prime - k) \rightarrow \frac{-1}{2p^\prime\cdot k}\,,
    \eeq
    which we write in the form
    \beq
    \Delta^{(\phi)}_F(p^\prime - k) \rightarrow \Delta_0(p^\prime,-k)\,,
    \eeq
    where we define
    \beq
    \label{defDelta0}
    \Delta_0(\ell_1,\ell_2) \equiv \frac{1}{2\ell_1\cdot\ell_2}\,.
    \eeq
    This gives
    \beqa
    \label{finalsimplifiedsigma12A}
    -i\left(\Sigma^{(A)}_{12}(k)\right)_{ba} & = &
    \int \Dpp (i\eta_1 g^\ast_b R)iS^{(f)}_{12}(p^\prime)(i\eta_2 g_a L)
    \nonumber\\
    &&\mbox{}\times
    \left(-i\Delta_0(p^\prime,-k)\right)
    \left(-i\pi^{(\phi)}_{21}(p^\prime - k)\right)i\Delta_0(p^\prime,-k)\,.
    \eeqa
    This is just the expression that we would obtain from \Eq{formalsigma12A}
    by considering only the term with $A = 1, B = 2$, and retain only
    the high $k$ limit of the vacuum part of the scalar propagator.

\end{enumerate}
Thus explicitly,
\beqa
\label{explicitsigma12Afinal}
-i\left(\Sigma^{(A)}_{12}(k)\right)_{ba} & = &
\sum_{c,d}\int\Dpp \Dp (i\eta_1 g^\ast_b R)iS^{(f)}_{12}(p^\prime)
(i\eta_2 g_a L)
(-i\Delta_{0}(p^\prime,-k))i\Delta_{0}(p^\prime,-k)
\nonumber\\
&&\mbox{} \times (-1)\left\{\mbox{Tr}(i\eta_1 g_c L)
\left(iS^{(\nu)}_{12}(k^\prime)\right)_{cd}
(i\eta_2 g^\ast_d R) iS^{(f)}_{21}(p)\right\}\,.
\eeqa

\subsection{Diagram B}

We now consider the contribution from diagram B in \Fig{fig:twoloop},
given in \Eq{explicitsigma12B}.
Here the pinch singularity involves the fermion propagator.
%
%
%\int\frac{d^4p}{(2\pi)^4}\frac{d^4p^\prime}{(2\pi)^4}
% 
We write it in the form
\beqa
\label{formalsigma12B}
-i\left(\Sigma^{(B)}_{12}(k)\right)_{ba} & = & \sum_{A,B}
\int\Dp i\Delta^{(\phi)}_{21}(p)(i\eta_1 g^\ast_b R)iS^{(f)}_{1A}(k + p)
\nonumber\\
&&\mbox{}\times
\left(-i\Sigma^{(f)}_{AB}(k + p)\right)
iS^{(f)}_{B2}(k + p) (i\eta_2 g_a L)\,,
\eeqa
where
\beq
\label{SigmaAB}
-i\Sigma^{(f)}_{AB}(k + p) = \sum_{c,d}
\int\Dpp i\Delta^{(\phi)}_{AB}(p^\prime)
(i\eta_A g_c L)\left(iS^{(\nu)}_{AB}(k^\prime)\right)_{cd}
(i\eta_B g^\ast_d R)\,.
\eeq
Therefore, in analogy with the previous case, here we denote by
$q = k + p$ the momentum of the virtual $f$, and consider the quantity
\beq
Y_{12}(q) \equiv \sum_{A,B} S^{(f)}_{1A}\Sigma^{(f)}_{AB}S^{(f)}_{B2} =
\left(\tilde S^{(f)}\tilde\Sigma^{(f)}\tilde S^{(f)}\right)_{12}\,,
\eeq
using the parametrization of the fermion propagator,
\beqa
\label{fermionparametrization}
\tilde S^{(f)} & = & U_f\left(
\begin{array}{cc}
  S^{(f)}_F & 0\\
  0 & -\bar S^{(f)}_F
\end{array}
\right)U_f
\,,\nonumber\\
\tilde\Sigma^{(f)} & = & U^{-1}_f\left(
\begin{array}{cc}
  \Sigma^{(f)} & 0\\
  0 & -\bar\Sigma^{(f)}
\end{array}
\right)U^{-1}_f\,,
\eeqa
where $S^{(f)}_F$ is the vacuum Feynman propagator.
The matrix $U_f$ is given by taking the expression given in \Eq{Uphi}
and replacing $\eta_\phi \rightarrow -\eta_f$, with $\eta_f$ defined
in \Eq{defeta}.

Therefore in correspondence with the scalar case here we have,
\beq
\tilde Y \equiv \tilde S^{(f)}\tilde\Sigma^{(f)}\tilde S^{(f)} =
U_f\left(
\begin{array}{cc}
  S^{(f)}_F\Sigma^{(f)}S^{(f)}_F & 0\\
  0 & -\bar S^{(f)}_F\bar\Sigma^{(f)}\bar S^{(f)}_F
\end{array}
\right)U_f\,.
\eeq
Thus as in the scalar case, the problematic terms $S^{(f)}_F\bar S^{(f)}_F$ that
give rise to the pinch singularity are actually absent. The quantity
that enters in the expression for $\Sigma^{(B)}_{12}$ is
\beqa
Y_{12} & = & U_{f\,11} U_{f\,12}\left[S^{(f)}_F\Sigma^{(f)}S^{(f)}_F -
  \bar S^{(f)}_F\bar\Sigma^{(f)}\bar S^{(f)}_F\right]
\nonumber\\
& = & -(\eta_f(q) - \theta(-q\cdot u))
\left[S^{(f)}_F\Sigma^{(f)}S^{(f)}_F -
  \bar S^{(f)}_F\bar\Sigma^{(f)}\bar S^{(f)}_F\right]\,.
\eeqa
Finally, it should be remembered that $\Sigma$ can be calculated
by using the formulas
\beqa
\label{Sigma}
\mbox{Re}\,\Sigma^{(f)} & \equiv &
\frac{1}{2}\left(\Sigma^{(f)} + \bar\Sigma^{(f)}\right) =
\mbox{Re}\,\Sigma^{(f)}_{11}\,,\nonumber\\
\mbox{Im}\,\Sigma^{(f)} & \equiv &
\frac{1}{2i}\left(\Sigma^{(f)} - \bar\Sigma^{(f)}\right) =
\frac{\Sigma^{(f)}_{12}}{2i[\eta_f(q) - \theta(-q\cdot u)]} =
\frac{\Sigma^{(f)}_{21}}{2i[\eta_f(q) - \theta(q\cdot u)]}\,,
\eeqa
where the $\Sigma^{(f)}_{AB}$ are given by the expressions in \Eq{SigmaAB}.

Thus in summary the procedure we need to follow is the following:
\begin{enumerate}
\item Compute $\mbox{Re}\,\Sigma^{(f)}_{11}(k + p)$ and
  $\Sigma^{(f)}_{12}(k + p)$ using \Eq{SigmaAB}, and from this
  determine $\Sigma^{(f)}(k + p)$ using \Eq{Sigma}.

\item Determine the quantity
  \beq
  \label{Y12}
  Y_{12}(q) = -(\eta_f(q) - \theta(-q\cdot u))
  \left[S^{(f)}_F\Sigma^{(f)}S^{(f)}_F -
    \bar S^{(f)}_F \bar\Sigma^{(f)}\bar S^{(f)}_F\right]\,.
  \eeq

  \item Compute $\Sigma^{(B)}_{12}(k)$ from
    \beqa
    \label{practicalsigma12B}
    -i\left(\Sigma^{(B)}_{12}(k)\right)_{ba} & = &
    \int\Dp i\Delta^{(\phi)}_{21}(p)(i\eta_1 g^\ast_b R)iY_{12}(p + k)
    (i\eta_2 g_a L)\,.
    \eeqa
    This is the result for diagram B, analogous to 
    \Eq{practicalsigma12A} for diagram A. It is free from the pinch
    singularities, and while it provides a consistent starting point,
    we can again simplify the explicit computation
    by proceeding as we did for diagram A.
      
  \item As we argue below, in the kinematic regime we are considering,
    the $Re\,\Sigma^{(f)}$ term in \Eq{Y12} contributes
    negligibly, so that we can take $Y_{12}$ to be
    \beq
    \label{Y12final}
    Y_{12} = -\frac{1}{2}\left[
      S^{(f)}_F\Sigma^{(f)}_{12}S^{(f)}_F +
      \bar S^{(f)}_F\Sigma^{(f)}_{12}\bar S^{(f)}_F
      \right]\,.
    \eeq
    The argument concerning the contribution from $\mbox{Re}\,\Sigma^{(f)}$
    is similar to the scalar case in the discussion of Diagram (A).
    Schematically, the term $S^{(f)\,2}_F - \bar S^{(f)\,2}_F$ contributes
    only when the $f$ is on-shell, that is
    $(k + p)^2 \approx m^2_f$. Since the external neutrino
    momentum $k$ as well as the scalar momentum $p$ are on shell
    in \Eq{practicalsigma12B}, the $\mbox{Re}\,\Sigma^{(f)}$ term corresponds
    to a contribution to $\Sigma^{(B)}_{12}$ arising from a process involving
    two-body subprocesses $\nu\phi \leftrightarrow f$. Again, under the
    kinematic conditions we are considering, such sub-processes are suppressed
    by the same (two-body) kinematics that suppress the one-loop contributions
    to $\Sigma^{(\nu)}_{12}$ (and the \emph{raison-de-entre} for considering the
    two-loop contributions) and therefore those terms can be neglected.
    
  \item More specifically, in our case in which we consider
    the high $k$ limit, we take
    \beq
    \label{defS0}
    S^{(f)}_F(k + p) \rightarrow S^{(f)}_0(p,k) \equiv
    \frac{\lslash{k} + \lslash{p} + m_f}{2p\cdot k}\,,
    \eeq
    which gives
    \beqa
    \label{finalsimplifiedsigma12B}
    -i\left(\Sigma^{(B)}_{12}(k)\right)_{ba} & = &
    \int\Dp i\Delta^{(\phi)}_{21}(p)(i\eta_1 g^\ast_b R)iS^{(f)}_{0}(p,k)
    \nonumber\\
    &&\mbox{}\times
    \left(-i\Sigma^{(f)}_{12}(p + k)\right)\left(-iS^{(f)}_0(p,k)\right)
    (i\eta_2 g_a L)\,.
    \eeqa
    This is just the expression that we would obtain from \Eq{formalsigma12B}
    by considering only the term with $A = 1, B = 2$, and retain only
    the high $k$ limit of the vacuum part of the $f$ propagator.

\end{enumerate}
Thus, explicitly,
\beqa
\label{explicitsigma12Bfinal}
-i\left(\Sigma^{(B)}_{12}(k)\right)_{ba} & = & 
\sum_{c,d}\int\Dp \Dpp 
i\Delta^{(\phi)}_{21}(p)i\Delta^{(\phi)}_{12}(p^\prime)
(i\eta_1 g^\ast_b R)iS^{(f)}_{0}(p,k)\nonumber\\
&&\mbox{}\times
(i\eta_1 g_c L)\left(iS^{(\nu)}_{12}(k^\prime)\right)_{cd}(i\eta_2 g^\ast_d R)
\left(-iS^{(f)}_{0}(p,k)\right) (i\eta_2 g_a L)\,.\nonumber\\
\eeqa

\subsection{Diagram C}

The contribution to the self-energy is
\beqa
-i\left(\Sigma^{(C)}_{12}(k)\right)_{ba}
& = & \sum_{A,B}\sum_{c,d}
\int\Dp \Dpp 
i\Delta^{(\phi)}_{2A}(p)i\Delta^{(\phi)}_{1B}(p^\prime)
(i\eta_1 g^\ast_b R)iS^{(f)}_{1A}(k - p^\prime) (i\eta_A g_c L)
\nonumber\\
&&\mbox{}\times
\left(iS^{(\nu)}_{AB}(k^\prime)\right)_{cd} (i\eta_B g^\ast_d R)
iS^{(f)}_{B2}(k + p) (i\eta_2 g_a L)\,,
\eeqa
Considering the sum over $A,B$, there are four terms, corresponding
to the combinations $AB = 12, 21, 11, 22$. Consider the third one,
that is $A = B = 1$. The integrand contains the factor
$S^{(f)}_{12}(k + p)\Delta^{(\phi)}_{21}(p)$, and as a consequence
of the delta functions involved, the momentum integration will be suppressed
by the same two-body kinematics we have already alluded to.
Similar arguments apply to the term $A = B = 2$, which contains the
factor $S^{(f)}_{12}(k - p^\prime)\Delta^{(\phi)}_{12}(p^\prime)$.

\begin{figure}
\begin{center}
\epsfig{file=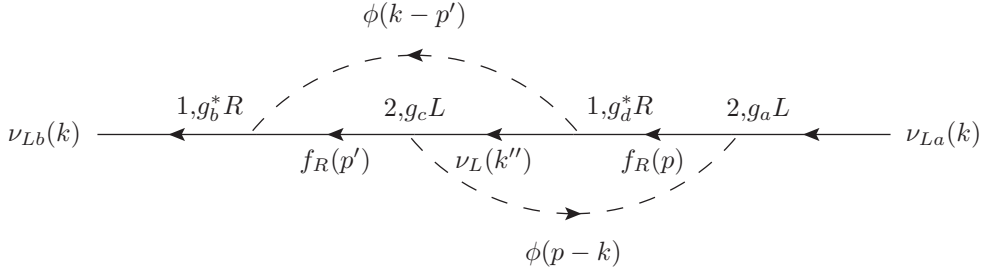,bbllx=123,bblly=361,bburx=489,bbury=461}
\end{center}
\caption[]{
  Equivalent form of diagram C in \Fig{fig:twoloop}, with a relabeling of
  the internal momentum variables. Here we have defined
  $k^{\prime\prime} = p + p^\prime - k$.
  \label{fig:diagramCeq}
}
\end{figure}

On the other hand, the terms corresponding to $AB = 12, 21$ are
not suppressed in this way. We therefore write
\beq
\left(\Sigma^{(C)}_{12}\right)_{ba} =
\left(\Sigma^{(CI)}_{12}\right)_{ba} +
\left(\Sigma^{(CII)}_{12}\right)_{ba}\,,
\eeq
where
\beqa
\label{explicitsigma12_CI}
-i\left(\Sigma^{(CI)}_{12}(k)\right)_{ba} 
& = & \sum_{c,d}
\int\Dp \Dpp 
i\Delta^{(\phi)}_{21}(p)i\Delta^{(\phi)}_{12}(p^\prime)
(i\eta_1 g^\ast_b R)iS^{(f)}_{11}(k - p^\prime)(i\eta_1 g_c L)
\nonumber\\
&&\mbox{}\times
\left(iS^{(\nu)}_{12}(k^\prime)\right)_{cd} (i\eta_2 g^\ast_d R)
iS^{(f)}_{22}(k + p) (i\eta_2 g_a L)\,,
\eeqa
and
\beqa
%\label{explicitsigma12_CII}
-i\left(\Sigma^{(CII)}_{12}(k)\right)_{ba}
& = & \sum_{c,d}
\int\Dp \Dpp 
i\Delta^{(\phi)}_{22}(p)i\Delta^{(\phi)}_{11}(p^\prime)
(i\eta_1 g^\ast_b R)iS^{(f)}_{12}(k - p^\prime)(i\eta_2 g_c L)
\nonumber\\
&&\mbox{}\times
\left(iS^{(\nu)}_{21}(k^\prime)\right)_{cd} (i\eta_1 g^\ast_d R)
iS^{(f)}_{12}(k + p) (i\eta_2 g_a L)\,.
\eeqa
For the purpose of carrying out the momentum integrations it is more
convenient to relabel the momentum variables in the expression
for $\left(\Sigma^{(CII)}_{12}\right)_{ba}$ according to the diagram
shown in \Fig{fig:diagramCeq}, which corresponds to
\beqa
\label{explicitsigma12_CII}
-i\left(\Sigma^{(CII)}_{12}(k\right)_{ba}
& = & \sum_{c,d}
\int\Dp \Dpp 
i\Delta^{(\phi)}_{22}(p - k)i\Delta^{(\phi)}_{11}(k - p^\prime)
(i\eta_1 g^\ast_b R)iS^{(f)}_{12}(p^\prime)(i\eta_2 g_c L)
\nonumber\\
&&\mbox{}\times
\left(iS^{(\nu)}_{21}(k^{\prime\prime})\right)_{cd} (i\eta_1 g^\ast_d R)
iS^{(f)}_{12}(p) (i\eta_2 g_a L)\,.
\eeqa

Invoking once again the two-body kinematics argument, we see that in
\Eq{explicitsigma12_CI} the contribution from the thermal part of the diagonal
$f$ propagators is suppressed, as well as the contribution from the
thermal part of the diagonal $\phi$ propagators in \Eq{explicitsigma12_CII}.
Thus, approximating the vacuum part of the diagonal propagators by
their high $k$ limit, 
\beqa
\label{explicitsigma12_CIfinal}
-i\left(\Sigma^{(CI)}_{12}(k)\right)_{ba}
& = & \sum_{c,d}
\int\Dp \Dpp 
i\Delta^{(\phi)}_{21}(p)i\Delta^{(\phi)}_{12}(p^\prime)
(i\eta_1 g^\ast_b R)iS^{(f)}_{0}(-p^\prime,k)(i\eta_1 g_c L)
\nonumber\\
&&\mbox{}\times
\left(iS^{(\nu)}_{12}(k^\prime)\right)_{cd} (i\eta_2 g^\ast_d R)
\left(-iS^{(f)}_{0}(p,k)\right) (i\eta_2 g_a L)\,,\\
\label{explicitsigma12_CIIfinal}
-i\left(\Sigma^{(CII)}_{12}(k)\right)_{ba} 
& = & \sum_{c,d}
\int\Dp \Dpp 
\left(-i\Delta^{(\phi)}_{0}(p,-k)\right)i\Delta^{(\phi)}_{0}(p^\prime,-k)
(i\eta_1 g^\ast_b R)iS^{(f)}_{12}(p^\prime)(i\eta_2 g_c L)
\nonumber\\
&&\mbox{}\times
\left(iS^{(\nu)}_{21}(k^{\prime\prime})\right)_{cd} (i\eta_1 g^\ast_d R)
iS^{(f)}_{12}(p) (i\eta_2 g_a L)\,.
\eeqa

\subsection{Summary}
\label{subsec-twoloop-pinch-summary}

In summary, we are left with four contributions,
\beq
\label{Sigma12reduced}
\left(\Sigma_{12}\right)_{ba} = \left(\Sigma^{(A)}_{12}\right)_{ba} +
\left(\Sigma^{(B)}_{12}\right)_{ba} +
\left(\Sigma^{(CI)}_{12}\right)_{ba} +
\left(\Sigma^{(CII)}_{12}\right)_{ba}\,,
\eeq
given by \Eqsss{explicitsigma12Afinal}{explicitsigma12Bfinal}
{explicitsigma12_CIfinal}{explicitsigma12_CIIfinal}.
In those formulas, $\Delta^{(\phi)}_0$ and $S^{(f)}_0$ are
the free fermion and scalar propagators, respectively, in the high
energy limit, defined in \Eqs{defDelta0}{defS0}.
From the expressions in those equations it follows that
only B and CI contribute in a pure $\phi$ background (no $f$ in the background)
while in a pure $f$ background (no $\phi$ in the background)
only A and CII contribute. In either case,
the diagrams CI and CII exist only if $\phi$ is a real scalar.
For a complex $\phi$ only diagrams A and B exist.
In the next section we consider precisely the former case, namely
a pure $\phi$ background. We determine the damping terms
according to the scheme explained in \Section{sec:preliminaries},
and evaluate explicitly the integrals involved
for some illustrative background conditions.

Before moving ahead, it is worth to emphasize the following.
Only diagrams A and B suffer from the pinch
singularities. \Eqs{explicitsigma12Afinal}{explicitsigma12Bfinal} 
give us convenient starting points to compute the contribution
from diagrams A and B, respectively, within the high-momentum
approximation that we restrict ourselves here.
However, the pinch singularities are already \emph{tamed}
in \Eqs{practicalsigma12A}{practicalsigma12B}, respectively,
and those expressions can be used to compute the corresponding contributions
in other situations of interest.

\section{A pure $\phi$ background}
\label{sec:purephibackground}

For definiteness, in the remainder of this work we consider only a $\phi$
background with no fermions $f$. We will calculate $\Sigma_i$ for
this case next. From the result we will determine the damping matrix
in \Section{subsec:purephibackgroundGamma}.

\subsection{Expression for $\Sigma_i$}
\label{subsec:purephibackgroundSigmai}

The starting point is the expression for each of the
diagrams B and CI we have obtained in
\Eqs{explicitsigma12Bfinal}{explicitsigma12_CIfinal}, explicitly
\beqa
\label{Sigma12_1}
-i\left(\Sigma^{(B)}_{12}(k)\right)_{ba} & = & (2K_{ba})
\int\frac{d^4p}{(2\pi)^4}\frac{d^4p^\prime}{(2\pi)^4}
i\Delta^{(\phi)}_{21}(p)i\Delta^{(\phi)}_{12}(p^\prime)\nonumber\\
&&\mbox{}\times
R iS^{(f)}_{0}(p,k) L iS^{(\nu)}_{12}(k^\prime)
R \left(-iS^{(f)}_{0}(p,k)\right)L\,,\nonumber\\
-i\left(\Sigma^{(CI)}_{12}(k)\right)_{ba} & = & (2K_{ba})
\int\frac{d^4p}{(2\pi)^4}\frac{d^4p^\prime}{(2\pi)^4}
i\Delta^{(\phi)}_{21}(p)i\Delta^{(\phi)}_{12}(p^\prime)\nonumber\\
&&\mbox{}\times
R iS^{(f)}_{0}(-p^\prime,k) L iS^{(\nu)}_{12}(k^\prime)
R \left(-iS^{(f)}_{0}(p,k)\right) L\,,
\eeqa
where
\beq
\label{Kba}
K_{ba} = \frac{1}{2} g^\ast_b g_a\left(\sum_c |g_c|^2\right)\,.
\eeq
We have used the fact that the neutrino propagator is diagonal
in flavor space, and we are defining $K_{ba}$ with the factor of $1/2$
for later convenience when we identify $\Sigma_i$.
It it understood that for a complex $\phi$ only $\Sigma^{(B)}_{12}$
contributes, while if $\phi$ is real then $\Sigma^{(CI)}_{12}$
also contributes but in that case the chemical potential $\alpha_\phi = 0$
in the final evaluation of the integrals involving the distribution
functions.

We write the formulas for the propagators as follows.
For the fermion $f$, remembering \Eq{defS0}, we have
\beq
RS^{(f)}_{0}(p,k)L = S_f(p,k)L\,,
\eeq
with
\beq
\label{Sf}
S_f(p,k) \equiv \frac{\lslash{p} + \lslash{k}}{2p\cdot k}\,.
\eeq
For the neutrino and the scalar propagators,
\beqa
\label{thermalpropagators}
S^{(\nu)}_{12}(k^\prime) & = & 2\pi i\delta(k^{\prime\,2})
n_F(x^\prime_\nu)\epsilon(k^\prime\cdot u) L s_\nu(k^\prime)\,,
\nonumber\\
\Delta^{(\phi)}_{21}(p) & = & -2\pi i\delta(p^2 - m^2_\phi)
e^{x_\phi} n_B(x_\phi)\epsilon(p\cdot u)\,,\nonumber\\
\Delta^{(\phi)}_{12}(p^\prime) & = & -2\pi i\delta(p^{\prime\,2} - m^2_\phi)
n_B(x^\prime_\phi)\epsilon(p^\prime\cdot u)\,,
\eeqa
where
\beq
s_\nu(k^\prime) = \lslash{k}^\prime\,,
\eeq
$\epsilon(z) = \theta(z) - \theta(-z)$ with $\theta(z)$ being the unit
step function, the fermion and boson distribution functions $n_{F,B}$ are
defined in \Eqs{nF}{nB}, and
\beqa
x^\prime_\nu & = & \beta k^\prime\cdot u - \alpha_\nu\,,\nonumber\\
x_\phi & = & \beta p\cdot u - \alpha_\phi \,,\nonumber\\
x^\prime_\phi & = & \beta p^\prime\cdot u - \alpha_\phi \,.
\eeqa
It is useful to remember that the relation $k^\prime = k + p - p^\prime$
(actually the definition of $k^\prime$), implies the following relation
\beq
x_\nu + x_\phi = x^\prime_\nu + x^\prime_\phi\,.
\eeq

Substituting \Eq{thermalpropagators} in \Eq{Sigma12_1}, we then have
\beqa
\label{Sigma12_2}
-i\left(\Sigma^{(B)}_{12}(k)\right)_{ba} & = & - (2K_{ba})
\int\frac{d^4p}{(2\pi)^3}\frac{d^4p^\prime}{(2\pi)^3}
\frac{d^4k^\prime}{(2\pi)^3}
(2\pi)^4\delta^{(4)}(k^\prime + p^\prime - k - p)
\nonumber\\
&&\mbox{}\times{}
\delta(p^2 - m^2_\phi)\delta(p^{\prime\,2} - m^2_\phi)\delta(k^{\prime\,2})
\epsilon(p\cdot u)\epsilon(p^\prime\cdot u)\epsilon(k^\prime\cdot u)
\nonumber\\
&&\mbox{}\times 
\left[S_f(p,k)s_\nu(k^\prime)S_f(p,k)L\right]E^\prime\,,\nonumber\\
-i\left(\Sigma^{(CI)}_{12}(k)\right)_{ba} & = & - (2K_{ba})
\int\frac{d^4p}{(2\pi)^3}\frac{d^4p^\prime}{(2\pi)^3}
\frac{d^4k^\prime}{(2\pi)^3}
(2\pi)^4\delta^{(4)}(k^\prime + p^\prime - k - p)\nonumber\\
&&\mbox{}\times
\delta(p^2 - m^2_\phi)\delta(p^{\prime\,2} - m^2_\phi)\delta(k^{\prime\,2})
\epsilon(p\cdot u)\epsilon(p^\prime\cdot u)\epsilon(k^\prime\cdot u)
\nonumber\\
&&\mbox{}\times 
\left[S_f(-p^\prime,k)s_\nu(k^\prime)S_f(p,k)L\right]E^\prime\,,
\eeqa
where
\beq
\label{Eprime}
E^\prime \equiv e^{x_\phi} n_B(x_\phi)n_B(x^\prime_\phi)n_F(x^\prime_\nu)\,.
\eeq
In writing \Eq{Sigma12_2} we have taken $k^\prime$ to be an arbitrary
variable but inserted a factor of $\delta^{(4)}(k + p - k^\prime - p^\prime)$
and integrate over $k^\prime$.

Letting $X = B,CI$, the corresponding contributions to the absorptive
part of the self-energy, identified by
\beq
\left(\Sigma_i\right)_{ba} = \frac{\Sigma_{12}}{2in_F(x_\nu)}\,,
\eeq
are then given by
\beqa
\label{Sigmai_1}
\left(\Sigma^{(X)}_{i}(k)\right)_{ba} & = & - K_{ba}
\int\frac{d^4p}{(2\pi)^3}\frac{d^4p^\prime}{(2\pi)^3}
\frac{d^4k^\prime}{(2\pi)^3}
(2\pi)^4\delta^{(4)}(k + p - k^\prime - p^\prime)
\nonumber\\
&&\mbox{}\times
\delta(p^2 - m^2_\phi)\delta(p^{\prime\,2} - m^2_\phi)\delta(k^{\prime\,2})
\epsilon(p\cdot u)\epsilon(p^\prime\cdot u)\epsilon(k^\prime\cdot u) 
M^{(X)}_\nu E\,,
\eeqa
where we have used the identity
\beq
\label{E}
\frac{1}{n_F(x_\nu)} E^\prime \equiv E =
n_B(x_\phi)\left(1 + n_B(x^\prime_\phi)\right) -
n_F(x^\prime_\nu)\left(n_B(x_\phi) - n_B(x^\prime_\phi)\right)\,,
\eeq
and we have defined
\beqa
\label{defMBCI}
M^{(B)}_\nu & = & S_f(p,k)s_\nu(k^\prime)S_f(p,k)L\,,\nonumber\\
M^{(CI)}_\nu & = & S_f(-p^\prime,k)s_\nu(k^\prime)S_f(p,k)L\,.
\eeqa
Next we carry out the integrals over $p^0, p^{\prime\,0}, k^{\prime\,0}$.
Starting with $k^{\prime\,0}$,
\beqa
\label{Sigmai_2}
\left(\Sigma^{(X)}_{i}(k)\right)_{ba} & = & - K_{ba}
\int\frac{d^4p}{(2\pi)^3}\frac{d^4p^\prime}{(2\pi)^3}
\frac{d^3\kappa^\prime}{(2\pi)^3 2\omega_{\kappa^\prime}}
\delta(p^2 - m^2_\phi)\delta(p^{\prime\,2} - m^2_\phi)
\epsilon(p\cdot u)\epsilon(p^\prime\cdot u)
\nonumber\\
&&\mbox{}\times (2\pi)^4\left\{
\delta^{(4)}(k + p - k^\prime - p^\prime)M^{(X)}_\nu E_\nu
- \delta^{(4)}(k + p + k^\prime - p^\prime)
M^{(X)}_{\bar\nu} E_{\bar\nu}\right\}\,,
\eeqa
where
\beqa
E_\nu & = & n_B(x_\phi)\left(1 + n_B(x^\prime_\phi)\right) -
f_\nu(\omega_{\kappa^\prime})\left(n_B(x_\phi) - n_B(x^\prime_\phi)\right)
\nonumber\\
E_{\bar\nu} & = & n_B(x^\prime_\phi)\left(1 + n_B(x_\phi)\right) +
f_{\bar\nu}(\omega_{\kappa^\prime})\left(n_B(x_\phi) - n_B(x^\prime_\phi)\right)
\,,
\eeqa
and
\beq
M^{(X)}_{\bar\nu} =
\left.M^{(X)}_{\nu}\right|_{k^\prime\rightarrow -k^\prime}\,,
\eeq
understanding that from now on
\beq
k^\prime = (\omega_{\kappa^\prime},\vec\kappa^\prime)\,,
\eeq
with $\omega_{\kappa^\prime} = |\vec\kappa^\prime| \equiv\kappa^\prime$.
To arrive at \Eq{Sigmai_2} we have also made the change of variable
$\vec\kappa^\prime \rightarrow -\vec\kappa^\prime$ in the second term.
Proceeding in a similar way with the integrals over $p^0$ and $p^{\prime\,0}$,
\beqa
\label{Sigmai_3}
\left(\Sigma^{(X)}_{i}(k)\right)_{ba} & = & - K_{ba}
\int\frac{d^3p}{(2\pi)^3 2E_p}\frac{d^3p^\prime}{(2\pi)^3 2E_{p^\prime}}
\frac{d^3\kappa^\prime}{(2\pi)^3 2\omega_{\kappa^\prime}}
\nonumber\\
&&\mbox{}\times \sum_{\lambda,\lambda^\prime} 
(2\pi)^4\left\{\delta^{(4)}(k + \lambda p - k^\prime - \lambda^\prime p^\prime)
M^{(X)}_{\nu,\lambda\lambda^\prime}
E_{\nu,\lambda\lambda^\prime}\right.\nonumber\\
&&\mbox{} - \left.
\delta^{(4)}(k + \lambda p + k^\prime - \lambda^\prime p^\prime)
M^{(X)}_{\bar\nu,\lambda\lambda^\prime}
E_{\bar\nu,\lambda\lambda^\prime}\right\}\,,
\eeqa
where
\beqa
\label{Mlambdalambdap}
M^{(B)}_{\nu,\lambda\lambda^\prime} & = &
S_f(\lambda p,k)s_\nu(k^\prime)S_f(\lambda p,k)L\,,\nonumber\\
M^{(CI)}_{\nu,\lambda\lambda^\prime} & = &
S_f(-\lambda^\prime p^\prime,k)s_\nu(k^\prime)S_f(\lambda p,k)L
\,,\nonumber\\
M^{(X)}_{\bar\nu,\lambda\lambda^\prime} & = &
\left.M^{(X)}_{\nu,\lambda\lambda^\prime}
\right|_{k^\prime\rightarrow -k^\prime}\,.
\eeqa
We have defined
\beqa
\label{Enu}
E_{\nu,\lambda\lambda^\prime} = \left.E_\nu\right|_{
p\rightarrow \lambda p,\,p^\prime \rightarrow \lambda^\prime p^\prime}
\eeqa
and similarly for $E_{\bar\nu,\lambda\lambda^\prime}$,
and from now on $p$ and $p^\prime$ are on-shell.
The formulas are given explicitly in \Table{table:processes}.
\begin{table}
\begin{center}
  \begin{tabular}{|l|l|}
    \hline
  $E_{\nu,++} = \fEp(1 + \fEpp) - \fnu(\fEp - \fEpp)$ &
  $\nu_{a,b}(k) + \phi(p) \leftrightarrow \nu_i(k^\prime) + \phi(p^\prime)$\\
  $E_{\nu,-+} = -(1 + \fbEp)(1 + \fEpp) + \fnu(1 + \fbEp + \fEpp)$ &
  $\nu_{a,b}(k) \leftrightarrow
  \nu_i(k^\prime) + \bar\phi(p) + \phi(p^\prime)$\\
  $E_{\nu,+-} = -\fEp\fbEpp - \fnu(1 + \fEp + \fbEpp)$ &
  $\nu_{a,b}(k) + \phi(p) + \bar\phi(p^\prime) \leftrightarrow
  \nu_i(k^\prime)$\\
  $E_{\nu,--} = (1 + \fbEp)\fbEpp + \fnu(\fbEp - \fbEpp)$ &    
  $\nu_{a,b}(k) + \bar\phi(p^\prime) \leftrightarrow
  \nu_i(k^\prime) + \bar\phi(p)$\\
  $E_{\bar\nu,++} = (1 + \fEp)\fEpp + \fnub(\fEp - \fEpp)$ &
  $\nu_{a,b}(k) + \bar\nu_i(\bar k^\prime) + \phi(p)\leftrightarrow
  \phi(p^\prime)$\\
  $E_{\bar\nu,-+} = -\fbEp\fEpp - \fnub(1 + \fbEp + \fEpp)$ &
  $\nu_{a,b}(k) + \bar\nu_i(\bar k^\prime) \leftrightarrow
  \bar\phi(p) + \phi(p^\prime)$\\
  $E_{\bar\nu,+-} = -(1 + \fEp)(1 + \fbEpp) + \fnub(1 + \fEp + \fbEpp)$ &
  $\nu_{a,b}(k) + \bar \nu_i(\bar k^\prime) +
  \phi(p) + \bar\phi(p^\prime) \leftrightarrow 0$\\
  $E_{\bar\nu,--} = \fbEp(1 + \fbEpp) - \fnub(\fbEp - \fbEpp)$ &
  $\nu_{a,b}(k) + \bar\nu_i(\bar k^\prime) + \bar\phi(p^\prime) \leftrightarrow
  \bar\phi(p)$\\
  \hline
\end{tabular}
\caption{Correspondence between the $E_{\nu,\lambda\lambda^\prime}$
  and $E_{\bar\nu,\lambda\lambda^\prime}$ factors defined in \Eq{Enu},
  and the process that contributes to the $\nu(k)$ damping via
  \Eq{Sigmai_3}. To simplify the notation we are using
    the shorthands shown in \Eq{fshorthand} for the various distribution
    functions.
  \label{table:processes}
}
\end{center}
\end{table}
To simplify the notation in the formulas summarized in \Table{table:processes}
we have introduce the shorthands
\beqa
\label{fshorthand}
f = f_\phi(E_p), \quad f^\prime = f_\phi(E_{p^\prime}), \quad
f^\prime_\nu = {f_\nu(\omega_{\kappa^\prime})}\nonumber\\
\bar f = f_{\bar\phi}(E_p), \quad \bar f^\prime = f_{\bar\phi}(E_{p^\prime}),
\quad
\bar f^\prime_\nu = f_{\bar\nu}(\omega_{\kappa^\prime})\,.
\eeqa
The formulas for $E_{\bar\nu,\lambda\lambda^\prime}$ are obtained from
those for $E_{\nu,\lambda\lambda^\prime}$ by making the replacement
$\fnu \rightarrow (1 - \fnub)$. Each of the terms
in \Eq{Sigmai_3} represents a contribution to $\Sigma_i$ from
a specific physical process, as indicated in \Table{table:processes}.

\subsection{Damping matrix}
\label{subsec:purephibackgroundGamma}

We write \Eq{defMBCI} in the form
\beq
\label{MMprimestart}
M^{(X)}_\nu = \frac{N^{(X)}L}{D^{(X)}}\,,\nonumber\\
\eeq
where
\beqa
D^{(B)} & = & (2p\cdot k)^2\,,\nonumber\\
D^{(CI)} & = & (-2p^\prime\cdot k)(2p\cdot k)\,,
\eeqa
and
\beqa
N^{(B)} & = &
(\lslash{k} + \lslash{p})\lslash{k}^\prime(\lslash{k} + \lslash{p})
\,,\nonumber\\
N^{(CI)} & = & (\lslash{k} - \lslash{p}^\prime)\lslash{k}^\prime(\lslash{k}
+ \lslash{p})\,.
\eeqa
The expressions for the $N^{(X)}$ are reduced by using the identity
\beq
\gamma_\lambda\gamma_\mu\gamma_\nu = C_{\lambda\mu\nu\rho}\gamma^\rho +
i\epsilon_{\lambda\mu\nu\rho}\gamma^\rho\gamma^5\,,
\eeq
where
\beq
C_{\lambda\mu\nu\rho} = g_{\lambda\mu}g_{\nu\rho} -
g_{\lambda\nu}g_{\mu\rho} + g_{\lambda\rho}g_{\mu\nu}\,.
\eeq
After the integrations over $k^\prime,p,p^\prime$ the only vectors
remaining are $k$ and $u$, and then the terms with
the antisymmetric tensor vanish. Thus we can replace in the integrand
\beq
N^{(X)} \rightarrow \ell^{(X)\rho}\gamma_\rho\,,
\eeq
where
\beqa
\ell^{(X)}_\rho & = & C_{\lambda\mu\nu\rho}
(k + p)^\lambda k^{\prime\,\mu}(k + p)^\nu\,,\nonumber\\
\ell^{(X)}_\rho & = & C_{\lambda\mu\nu\rho}
(k - p^\prime)^\lambda k^{\prime\,\mu}(k + p)^\nu\,.
\eeqa
Then, corresponding to each diagram $X = B,CI$ we have
\beqa
\label{Vi_1}
\left(V^{(X)\mu}_{i}(\omega,\vec\kappa)\right)_{ba} & = & - K_{ba}
\int\frac{d^3p}{(2\pi)^3 2E_p}\frac{d^3p^\prime}{(2\pi)^3 2E_{p^\prime}}
\frac{d^3\kappa^\prime}{(2\pi)^3 2\omega_{\kappa^\prime}}
\nonumber\\
&&\mbox{}\times \sum_{\lambda,\lambda^\prime}
\left(
\frac{\ell^{(X)\mu}_{\lambda\lambda^\prime}}{D^{(X)}_{\lambda\lambda^\prime}}
\right)
(2\pi)^4\left\{\delta^{(4)}(k + \lambda p - k^\prime - \lambda^\prime p^\prime)
E_{\nu,\lambda\lambda^\prime}\right.\nonumber\\
&&\mbox{} + \left.
\delta^{(4)}(k + \lambda p + k^\prime - \lambda^\prime p^\prime)
E_{\bar\nu\lambda\lambda^\prime}\right\}\,,
\eeqa
where
\beqa
\ell^{(X)\mu}_{\lambda\lambda^\prime} & \equiv &
\left.\ell^{(X)\mu}\right|_{p\rightarrow\lambda p,
  p^\prime\rightarrow\lambda^\prime p^\prime},\nonumber\\ 
D^{(X)}_{\lambda\lambda^\prime} & \equiv &
\left.D^{(X)}\right|_{p\rightarrow\lambda p,
  p^\prime\rightarrow\lambda^\prime p^\prime}\,.
\eeqa

Since the formulas for $\Gamma$ are given in terms of
$n\cdot V_i(\kappa,\vec\kappa)$ (for neutrinos), or
$n\cdot V_i(-\kappa,-\vec\kappa)$ (for antineutrinos),
we consider the evaluation of $\Sigma_i$ for
$k^\mu = \omega n^\mu$ (which in particular implies $k^2 = 0$),
and in the end put $\omega = \pm \kappa$ (for neutrinos
or antineutrinos, respectively). From now on we thus set $k^2 = 0$.

We evaluate $n\cdot\ell$ and $n\cdot\ell^\prime$,
putting $k^\mu = \omega n^\mu$ as we already stated.
Then doing the algebra, remembering to set $n^2 = 0$,
\beqa
n\cdot \ell^{(B)} & = & 2(n\cdot p)(k^\prime\cdot p)\,,\nonumber\\
n\cdot \ell^{(CI)} & = & -(n\cdot p)(k^\prime\cdot p^\prime) -
(n\cdot p^\prime)(k^\prime\cdot p)\,,
\eeqa
and therefore
\beqa
\frac{n\cdot\ell^{(B)}}{D^{(B)}} & = &
\frac{1}{2\omega}\frac{k^\prime\cdot p}{k\cdot p}\,,\nonumber\\
\frac{n\cdot\ell^{(CI)}}{D^{(CI)}} & = & \frac{1}{4\omega}
\frac{k^\prime\cdot p^\prime}{k\cdot p^\prime} +
\frac{1}{4\omega}\frac{k^\prime\cdot p}{k\cdot p}\,.
\eeqa
Up to this moment we have only used straightforward algebra to arrive
here from \Eq{MMprimestart}. We now invoke the high energy limit we are
considering. The momentum delta functions set $k^\prime = k + p - p^\prime$.
Therefore, to leading order in $k$, we put $k^\prime\rightarrow k$ in the above
and for either diagram we have
\beq
\frac{n\cdot\ell^{(X)}}{D^{(X)}} = \frac{1}{2\omega}\,.
\eeq
For the antineutrino part, the delta function gives
$k^\prime = - k - p + p^\prime$, therefore the replacement
is $k^\prime \rightarrow -k$.
Putting all this together we then have, from \Eq{Vi_1},
\beqa
\label{nVi}
\left(n\cdot V^{(X)}_i(\omega,\omega\hat\kappa)\right)_{ba} & = &
-\frac{K_{ba}}{2\omega}
\int\frac{d^3p}{(2\pi)^3 2E_p}\frac{d^3p^\prime}{(2\pi)^3 2E_{p^\prime}}
\frac{d^3\kappa^\prime}{(2\pi)^3 2\omega_{\kappa^\prime}}
\nonumber\\
&&\mbox{}\times \sum_{\lambda,\lambda^\prime} 
(2\pi)^4\left\{\delta^{(4)}(k + \lambda p - k^\prime - \lambda^\prime p^\prime)
E_{\nu,\lambda\lambda^\prime}\right.\nonumber\\
&&\mbox{} - \left.
\delta^{(4)}(k + \lambda p + k^\prime - \lambda^\prime p^\prime)
E_{\bar\nu\lambda\lambda^\prime}\right\}\,.
\eeqa
The damping matrix is given by
\beq
\label{GammafinalformulaB}
-\frac{1}{2}\Gamma^{(2)} = \left\{\begin{array}{ll}
n\cdot V^{(B)}_i(\kappa,\vec\kappa) & (\nu)\\
n\cdot V^{(B)\ast}_i(-\kappa,-\vec\kappa) & (\bar\nu)\,,
\end{array}\right.
\eeq
or
\beq
\label{GammafinalformulaCI}
-\frac{1}{2}\Gamma^{(2)} = \left\{\begin{array}{ll}
n\cdot V^{(B)}_i(\kappa,\vec\kappa) +
n\cdot V^{(CI)}_i(\kappa,\vec\kappa) & (\nu)\\
n\cdot V^{(B)\ast}_i(-\kappa,-\vec\kappa) +
n\cdot V^{(CI)\ast}_i(-\kappa,-\vec\kappa) & (\bar\nu)\,,
\end{array}\right.
\eeq
for a complex or real $\phi$, respectively. The final expressions
for both diagram contributions $n\cdot V^{(X)}_i$, given in \Eq{nVi},
are formally the same. But it must be understood that for a real $\phi$
the distribution functions of the $\phi$ have $\alpha_\phi = 0$,
or equivalently $f_{\bar\phi} = f_{\phi}$.

Not all the terms in \Eq{nVi} contribute, depending on whether $\omega$
is positive or negative. Equivalently, the corresponding processes
are inhibited by the kinematics. In addition we will assume that
there are no neutrinos in the background. The result is that
for the neutrinos ($\omega$ positive) only the terms $E_{\nu,++},E_{\nu,--}$
contribute, while for the antineutrinos ($\omega$ negative) only
$E_{\bar\nu,++},E_{\bar\nu,--}$ contribute.

Denoting by $\Gamma^{(\nu)}$ and $\Gamma^{(\bar\nu)}$ the matrices
for neutrinos and antineutrinos, respectively, for the case
of a complex $\phi$ we then have from \Eq{nVi},
\beqa
\label{Gammaexplicitnubarnu}
\frac{1}{2}\Gamma^{(\nu)}_{ba} & = &
\frac{K_{ba}\gamma^{(\nu)}}{2\kappa}\,,\nonumber\\
\frac{1}{2}\Gamma^{(\bar\nu)}_{ba} & = &
\frac{K^\ast_{ba}\gamma^{(\bar\nu)}}{2\kappa}\,,
\eeqa
where
\beqa
\label{gammalinblad}
\gamma^{(\nu)} & = &
\int\frac{d^3p}{(2\pi)^3 2E_p}\frac{d^3p^\prime}{(2\pi)^3 2E_{p^\prime}}
\frac{d^3\kappa^\prime}{(2\pi)^3 2\omega_{\kappa^\prime}}
\nonumber\\
&&\mbox{} (2\pi)^4\left\{\delta^{(4)}(k + p - k^\prime - p^\prime) E_{\nu,++} +
\delta^{(4)}(k + p^\prime - k^\prime - p) E_{\nu,--}\right\}\,,\nonumber\\
\gamma^{(\bar\nu)} & = &
\int\frac{d^3p}{(2\pi)^3 2E_p}\frac{d^3p^\prime}{(2\pi)^3 2E_{p^\prime}}
\frac{d^3\kappa^\prime}{(2\pi)^3 2\omega_{\kappa^\prime}}
\nonumber\\
&&\mbox{}
(2\pi)^4\left\{\delta^{(4)}(k + p^\prime - k^\prime - p) E_{\bar\nu,++} +
\delta^{(4)}(k + p - k^\prime - p^\prime) E_{\bar\nu,--}\right\}\,,
\eeqa
with
\beqa
E_{\nu ++} & = & f_\phi(E_p)(1 + f_\phi(E_{p^\prime}))\,,\nonumber\\
E_{\nu --} & = & f_{\bar\phi}(E_{p^\prime})(1 + f_{\bar\phi}(E_{p}))
\,,\nonumber\\
E_{\bar\nu ++} & = & f_\phi(E_{p^\prime})(1 + f_\phi(E_{p}))\,,\nonumber\\
E_{\bar\nu --} & = & f_{\bar\phi}(E_{p})(1 + f_{\bar\phi}(E_{p^\prime}))\,.
\eeqa
Relabeling the $p,p^\prime$ integration variables
in some terms, we can see that $\gamma^{(\bar\nu)} = \gamma^{(\nu)}$.
Therefore, in explicit form, \Eq{Gammaexplicitnubarnu} becomes
\beqa
\label{Gammaexplicit}
\frac{1}{2}\Gamma^{(\nu)}_{ba} & = &
\frac{K_{ba}\gamma^{(\phi)}}{2\kappa}\,,\nonumber\\
\frac{1}{2}\Gamma^{(\bar\nu)}_{ba} & = &
\frac{K^\ast_{ba}\gamma^{(\phi)}}{2\kappa}\,,
\eeqa
where
\beqa
\label{defgammaphi}
\gamma_\phi & \equiv &
\int\frac{d^3p}{(2\pi)^3 2E_p}\frac{d^3p^\prime}{(2\pi)^3 2E_{p^\prime}}
\frac{d^3\kappa^\prime}{(2\pi)^3 2\omega_{\kappa^\prime}}
\nonumber\\
&&\mbox{} (2\pi)^4\delta^{(4)}(k + p - k^\prime - p^\prime)
\left\{f_\phi(E_p)(1 + f_\phi(E_{p^\prime})) +
f_{\bar\phi}(E_p)(1 + f_{\bar\phi}(E_{p^\prime}))\right\}\,.
\eeqa
As already stated, \Eq{Gammaexplicit} holds for a complex $\phi$.
For a real $\phi$ the formula is the same
but with the replacement $2\kappa \rightarrow \kappa$,
and putting $f_{\bar\phi} = f_\phi$ in \Eq{defgammaphi}.

\subsection{Example evaluation of integrals}

In the dilute gas approximation (i.e., neglecting the terms
with the product of the distribution function),
\beq
\gamma_\phi = \int\frac{d^3p}{(2\pi)^3 2E_p}
(f_\phi(E_p) + f_{\bar\phi}(E_p))J\,,
\eeq
where
\beq
J = \int\frac{d^3p^\prime}{(2\pi)^3 2E_{p^\prime}}
\frac{d^3\kappa^\prime}{(2\pi)^3 2\omega_{\kappa^\prime}}
(2\pi)^4\delta^{(4)}(k + p - k^\prime - p^\prime)\,.
\eeq
It is straightforward to evaluate the $J$ integral. Let us define
\beq
q = k + p\,,
\eeq
and
\beq
\label{s}
s = q^2\,.
\eeq
We then obtain
\beq
\label{Jfinal}
J = \frac{1}{4\pi}\left(\frac{p^{\prime\,\ast}}{\sqrt{s}}\right)
\theta(\sqrt{s} - E^{\ast}_{p^\prime})\theta(E^{\ast}_{p^\prime} - m_\phi)\,,
\eeq
where
\beq
E^{\ast}_{p^\prime} = \frac{s + m^2_\phi}{2\sqrt{s}}\,,
\eeq
and
\beq
p^{\prime\,\ast} = \sqrt{E^{\ast\,2}_{p^\prime} - m^2_\phi} =
\frac{s - m^2_\phi}{2\sqrt{s}}\,.
\eeq
From the definition in \Eq{s},
\beqa
s & = & m^2_\phi + 2k\cdot p\nonumber\\
& = & m^2_\phi + 2\kappa E_p(1 - v_p\cos\theta_p)\,, 
\eeqa
where
\beq
\cos\theta_p = \hat\kappa\cdot\hat p\,,
\eeq
and
\beq
v_p = \frac{|\vec p|}{E_p}\,.
\eeq
Thus for any value of $\vec p$, we have $s > m^2_\phi$, and this
implies that the two step functions in \Eq{Jfinal} are automatically
satisfied. Therefore, we can take $J$ to be simply
\beq
J = \frac{1}{8\pi}\left(\frac{s - m^2_\phi}{s}\right) =
\frac{1}{8\pi}\left(1 - \frac{m^2_\phi}{s}\right) = 
\frac{1}{8\pi}\left(1 - \frac{m^2_\phi}
     {m^2_\phi + \kappa E_p(1 - v_p\cos\theta_p)}\right)
\eeq

In principle we can use this to do the remaining integral over $\vec p$
to evaluate $\gamma_{\phi}$ for different background distribution functions.
However, since we are interested in
the high $\kappa$ limit, we retain just the leading term
\beq
J = \frac{1}{8\pi}\,,
\eeq
which in turn gives
\beq
\label{gammaphi}
\gamma_\phi = \frac{1}{8\pi} J_\phi\,.
\eeq
Thus from \Eq{Gammaexplicit}, for a complex $\phi$,
\beq
\label{Gammaexplicit1}
\frac{1}{2}\Gamma^{(\nu)}_{ba} = \frac{K_{ba}J_\phi}{16\pi\kappa}\,,
\eeq
where $J_\phi$ is defined in \Eq{Jfphi}. For a real $\phi$
\beq
\label{Gammaexplicit1real}
\frac{1}{2}\Gamma^{(\nu)}_{ba} = \frac{K_{ba}J_\phi}{8\pi\kappa}\,,
\eeq
and $J_\phi$ is evaluated putting $f_{\bar\phi} = f_\phi$.
In either case the formula for $\Gamma^{(\nu)}_{ba}$ is obtained
by replacing $K_{ba}\rightarrow K^\ast_{ba}$. Thus, for example,
using \Eq{Jphinr-er}, \Eq{Gammaexplicit1} yields
\beq
\label{Gammaexplicit2}
\frac{1}{2}\Gamma^{(\nu)}_{ba} = K_{ba}\left\{
\begin{array}{ll}
  \frac{n_\phi + n_{\bar\phi}}{32\pi\kappa m_\phi}
  & \mbox{(NR)} \\[12pt]
  \frac{T^2}{192\pi\kappa} & \mbox{(UR)}\,.
  \end{array}\right.
\eeq
These are valid for a complex $\phi$. For a real $\phi$,
the corresponding formulas are
\beq
\label{Gammaexplicit2real}
\frac{1}{2}\Gamma^{(\nu)}_{ba} = K_{ba}\left\{
\begin{array}{ll}
  \frac{n_\phi}{8\pi\kappa m_\phi}
  & \mbox{(NR)} \\[12pt]
  \frac{T^2}{96\pi\kappa} & \mbox{(UR)}\,.
  \end{array}\right.
\eeq
For the antineutrinos, the damping matrix is given by the same
formulas, but replacing $K_{ba} \rightarrow K^\ast_{ba}$.

Comparing \Eq{Gammaexplicit1} (or \Eq{Gammaexplicit1real})
with \Eq{Vfphi} we see that the ratio of the imaginary part (damping)
to the real part of the effective potential is $\sim g^2/16\pi$. This
contrasts with the result in the case of a normal matter background.
In that case the same ratio is further suppressed by the mass factor
$\kappa m_e/m^2_W$ (for $\kappa > m_e$) or $\kappa^2/m^2_W$
(for $m_e > \kappa$)\cite{nsnudecsm}.
Therefore, in situations where the effective potential due to a
light scalar background may be relevant, the damping effects
may be important since they are not suppressed by the mass factors.
On the other hand, the relative importance of such
damping effects may be negligible if all $g_a$ couplings are too small.

\subsection{Discussion}
\label{subsec:calculation-damping-discussion}

We have obtained \Eqs{Gammaexplicit2}{Gammaexplicit2real}, or their more
general versions given in \Eqs{Gammaexplicit1}{Gammaexplicit1real},
by purposely considering a background with only
$\phi$ particles and no fermions $f$. However the inclusion of the fermion
contribution can be carried out straightforwardly in analogous fashion.
It involves calculating in a similar way the contributions
denoted by $\Sigma^{(A)}_{12}$ and $\Sigma^{(CII)}_{12}$ in \Eq{Sigma12reduced}.

From a physical point of view, the damping matrix $\Gamma$ induces
decoherence effects in the propagation of neutrinos.
As emphasized in our previous work \xRef{nsnuphidec}, and illustrated again here,
the contribution to $\Gamma$ from the neutrino non-forward scattering process
$\nu_a + x \rightarrow \nu_b + x$, where $x = f,\phi$, can be determined
from the two-loop calculation of $\Sigma_i$.
However, since in this case the initial neutrino state is depleted
but the neutrino does not actually disappear (the initial neutrino
transitions into a neutrino of a different flavor but does not decay
into a $f\phi$ pair, for example), we have argued that the effects of
the non-forward scattering processes are more appropriately interpreted
in terms of decoherence phenomena rather than damping.
Specifically, the damping matrix should be associated with decoherence
effects in terms of the Lindblad equation
and the notion of the stochastic evolution of the state
vector\cite{Daley:2014fha,Weinberg:2011jg, pearle,Plenio:1997ep,Lieu:2019cev}.
The idea is to assume that the evolution due to the
damping effects described by $\Gamma$ is accompanied by a stochastic evolution
that cannot be described by the coherent evolution of the state vector.
As discussed in detail in \xRef{nsnuphidec}, the result of this idea is that
the evolution of the system is described by the density
matrix $\rho$ (in the sense that we can use it to calculate averages of quantum
expectation values) that satisfies the Lindblad equation,
\beq
\label{lindbladeq}
\partial_t\rho = -i[H_r,\rho] + \sum_n
\left\{L_n \rho L^\dagger_n - \frac{1}{2}L^\dagger_n L_n\rho -
\frac{1}{2}\rho L^\dagger_n L_n\right\}\,,
\eeq
where the $L_n$ matrices, representing the \emph{jump operators},
are related to $\Gamma$ by
\beq
\label{defL}
\Gamma = \sum_n L^\dagger_n L_n\,.
\eeq
We refer to the terms involving the jump operators in the
right-hand-side of \Eq{lindbladeq} as the decoherence terms.

The damping matrix that we have determined from the two-loop self-energy
calculation can be expressed in this form. For example, consider a real
$\phi$ background. \Eq{Gammaexplicit1real} can be written as
\beq
\Gamma^{(\ell)}_{ba} = (L^{(\ell)\dagger}_\phi L^{(\ell)}_\phi)_{ba} =
\sum_c (L^{(\ell)}_\phi)^\ast_{cb}
(L^{(\ell)}_\phi)_{ca}  \qquad (\ell = \nu,\bar\nu)\,,
\eeq
with
\beqa
\label{L1}
(L^{(\nu)}_\phi)_{ca} & = & \sqrt{\frac{J_\phi}{8\pi\kappa}}\,
g_c g_a\,,\nonumber\\
(L^{(\bar\nu)}_\phi)_{ca} & = & (L^{(\nu)}_\phi)^\ast_{ca}\,.
\eeqa
The $L$ matrices are expressed in terms of integrals over the background
particles distribution functions. Going a step further, consider for
illustrative purposes the NR limit. Then using \Eq{Jphinr-real},
\beqa
\label{L1nrer}
(L^{(\nu)}_\phi)_{ca} & = & \sqrt{\frac{n_\phi}{8\pi\kappa m_\phi}}\,
g_c g_a\,,\nonumber\\
(L^{(\bar\nu)}_\phi)_{ca} & = & \sqrt{\frac{n_\phi}{8\pi\kappa m_\phi}}\,
g^\ast_c g^\ast_a\,.
\eeqa

It is straightforward to consider the addition of fermions $f$ in
the background, or in fact more complicated superpositions of different
background species. The evaluation of $\Sigma^{(A,CII)}_{12}$ would result in a
matrix $L_f$ contributing in \Eq{defL}, so that
\beq
\Gamma^{(\ell)} = L^{(\ell)\dagger}_\phi L^{(\ell)}_\phi +
L^{(\ell)\dagger}_f L^{(\ell)}_f\,.
\eeq
The matrix $L^{(\ell)}_f$ would be given in terms of the fermion
distribution by formulas analogous to \Eq{L1}.
In general these formulas predict, for example,
a specific dependence of the decoherence terms on the neutrino energy,
depending on the background conditions. This complements the
studies of the decoherence effects that are based on general
considerations at a phenomenological level without a calculation
of the decoherence terms. We do not pursue this any further here,
but our results and calculations show the path
for further applications along these lines.

\section{Conclusions and outlook}
\label{sec:conclusions}

In this work we have been concerned with the calculation of the damping terms
that result from non-forward scattering processes when neutrinos propagate
in a background of fermions ($f$) and scalars ($\phi$) interacting
via a Yukawa-type interaction. We determine the
contribution of those processes to the damping matrix $\Gamma$
from the two-loop calculation of the imaginary part of the
thermal neutrino self-energy using the methods of thermal field theory (TFT).
In the context of TFT the two-loop self-energy diagrams suffer
from the so-called \emph{pinch singularities}, which appear because
the expressions contain two thermal propagators with the same momentum
argument. A significant effort in this work was to show how
those singularities are effectively handled by a 
judicious use of the properties and parametrizations of the
thermal propagators. The final result of that exercise is a set
of formulas for the two-loop contribution to the imaginary part
of the self-energy from which the damping matrix
is determined. The formulas are well-defined integrals over
the background particle momentum distribution functions, which
can be evaluated straightforwardly for different background conditions.
For concreteness, we considered in detail a pure $\phi$ background,
with no fermions $f$. We obtained the corresponding formulas
for the damping terms, and evaluated them in some specific limits
of the $\phi$ distribution functions.

As a guide to applications, we discussed briefly in
\Section{subsec:calculation-damping-discussion} the connection
between $\Gamma$ and the decoherence described in terms
of the Lindblad equation. There we showed the explicit formulas
obtained for the \emph{jump operators} that appear in the Lindblad equation
using the results of the calculation of $\Gamma$.
We indicated how this approach can be extended to
consider more general backgrounds that include the fermions $f$
or other particles.

The results we have presented extend our previous work
and can be used to study the decoherence effects in a variety of physical
contexts and environments. As a by-product, we have presented a
detailed calculation of the imaginary part of the two-loop neutrino thermal
self-energy, controlling the pinch singularities,
using and illustrating a method that can be applied
to perform similar calculations consistently in other situations of interest.

%\begin{acknowledgments}
The work of S. S. is partially supported by DGAPA-UNAM
(Mexico) Project No. IN103019.
%\end{acknowledgments}

%\bibliographystyle{ieeetr}
%\bibliography{main}

\end{document}